# OPERADIC FORMULATION OF TOPOLOGICAL VERTEX ALGEBRAS AND GERSTENHABER OR BATALIN-VILKOVISKY ALGEBRAS

YI-ZHI HUANG

Abstract. We give the operadic formulation of (weak, strong) topological vertex algebras, which are variants of topological vertex operator algebras studied recently by Lian and Zuckerman. As an application, we obtain a conceptual and geometric construction of the Batalin-Vilkovisky algebraic structure (or the Gerstenhaber algebra structure) on the cohomology of a topological vertex algebra (or of a weak topological vertex algebra) by combining this operadic formulation with a theorem of Getzler (or of Cohen) which formulates Batalin-Vilkovisky algebras (or Gerstenhaber algebras) in terms of the homology of the framed little disk operad (or of the little disk operad).

## 1. Introduction

Recently operads have received a lot of attentions from mathematicians working in different areas. Many complicated algebraic structures can now be formulated and studied conceptually using the language of operads. In the present paper, we give the operadic formulation of another type of algebraic structures — (weak, strong) topological vertex algebras — which are variants of topological vertex operator algebras defined in [LZ]. As an application, we obtain a geometric construction of the Batalin-Vilkovisky algebraic structure (or the Gerstenhaber algebra structure) on the cohomology of a topological vertex algebra (or of a weak topological vertex algebra) using this operadic formulation and a recent theorem of Getzler [Get] (or a theorem of Cohen [C] [Get]).

Operads are devices to describe operations. For classical algebraic structures, the correponding operads are so simple (geometrically they are usually constructed from one-dimensional objects) that these operads themselves do not have any interesting structure to be studied. Even though the notion of operad is very natural, it would be only a fancy language rather than a necessary and deep way to understand operations conceptually for these classical algebraic structures. But the situation changes when

The author is supported in part by NSF grant DMS-9104519.

we consider more complicated operations. The first important example of an operad-like structure occurred in Stasheff's notion of $A_\infty$-space [S1]. Unlike the operads for classical algebraic structures, it has rich geometric structure. In [May], May formalized the notion of operad and began to use it systematically in the study of iterated loop spaces. The recent studies of Gerstenhaber algebras, Batalin-Vilkovisky algebras, homotopy Lie algebras and vertex operator algebras in terms of operads (see, for example, [Get] [HS] [HL1] [HL2]) show the power and importance of the notion of operad. See also [S2] for a recent survey.

The notion of vertex operator algebra and its various variants arises naturally in the vertex operator construction of the Monster sporadic group by Frenkel, Lepowsky and Meurman [FLM1]. It is essentially the same as the notion of chiral algebra in conformal field theory (see, e.g., [MS]). In [B] Borcherds introduces the notion of vertex algebra based on his insightful understanding of [FLM1]. In [FLM2], the notion of vertex operator algebra — a modification of that of vertex algebra — is introduced and it is proved that the moonshine module constructed in [FLM1] has the structure of a vertex operator algebra such that the Monster is its automorphism group. In [FHL], the basic axiomatic theory of vertex operator algebras is developed. The difference between vertex algebras and vertex operator algebras is that vertex algebras do not have to satisfy those axioms on the grading and do not have to have a Virasoro element. We can also define some other variants of vertex operator algebras by relaxing or generalizing some other axioms. For various examples, variants and generalizations of vertex operator algebras, see, for example, [FLM2] [FHL] [T] [FFR] [H3] [FZ] [DL] [L].

In [H1] [H2] [HL1] [HL2] [H4], it has been established that vertex operator algebras are in fact algebras over the partial operads of powers of the determinant line bundle over the partial operad $K$ of a certain moduli space of spheres with punctures and local coordinates, satisfying a certain meromorphicity axiom. A small part of this operadic interpretation of vertex operator algebras in fact shows that vertex algebras are algebras over a partial suboperad $\hat{K}$ of $K$ satisfying the part of meromorphicity axiom which still make sense for $\hat{K}$. In this paper, the algebraic structures we will study are variants of topological vertex operator algebra introduced by Lian and Zuckerman: weak topological vertex algebras, topological vertex algebras and strong topological vertex algebras (see Definition 2.1). The main purpose of the present paper is to give a completely geometric formulation of these topological vertex algebras using the structure of the partial operad $\hat{K}$.

We show that the category of topological vertex algebras (or weak topological vertex algebras) is isomorphic to a category whose objects are triples of the form $(V, Q, \{\omega_j | j \in \mathbb{N}\})$ where $V$ is a $\mathbb{Z} \times \mathbb{Z}$-graded vector space, $Q$ an operator on $V$ of degree 1 with respect to the second $\mathbb{Z}$-grading and $\omega_j$ a holomorphic form on $\hat{K}(j)$ for every $j$ in the set $\mathbb{N}$ of nonnegative integers, satisfying certain natural axioms. An immediate consequence is that the operad of the homology of the framed little disk operad (or the little disk operad) acts on the cohomology of a topological vertex algebra (or of a weak topological vertex algebra). Combined with Getzler's theorem (or Cohen's theorem) which describes Batalin-Vilkovisky algebras (or Gerstenhaber algebras) in terms of the operad of the homology of the framed little disk operad (or of the little disk operad), it gives the Batalin-Vilkovisky algebra structure (or the Gerstenhaber structure) on the cohomology of a topological vertex algebra (or of a weak topological vertex algebra). This last result has been obtained first by Lian and Zuckerman [LZ]. Also, beginning with the geometric axioms of topological conformal field theories, Getzler has shown that the cohomology of a topological conformal field theory has the structure of a Batalin-Vilkovisky algebra [Get]. The main new result of the present paper is the equivalence between the algebraic formulation and the geometric formulation of (weak, strong) topological vertex algebras and, as a consequence, Lian-Zuckerman's result can now be understood conceptually using the results of this paper and Getzler's theorem or Cohen's theorem. This equivalence also provides examples of what Getzler begins with. In fact it can be shown further that the category of strong topological vertex algebras is isomorphic to the category of algebras over the partial operad of the differentiable chain complexes of the partial operad $\hat{K}$ satisfying a certain meromorphicity axiom. But since the meromophicity axiom for these algebras over the differentiable chain complexes is much more complicated than that for $\omega_j$, $j \in \mathbb{N}$ and the formulation in terms of these forms is already conceptual enough, we only give half of this further formulation for future topological applications; we show that these forms give a morphism of differential graded partial operads from the partial operad of the differentiable chain complexes of the partial operad $\hat{K}$ to the endomorphism partial pseudo-operad for $V$.

One remark on the algebraic and geometric formulations of quantum field theories might be helpful here. In general, the geometric formulations of quantum field theories are axiomizations of the path integral approach in physics and have the advantage that they give conceptually satisfactory definitions and they also allow one to derive many important results using the geometric intuition. One famous example is the Verlinde formula. In [MS], it is shown at the physical level of rigor that the geometric axioms plus the rationality imply the Verlinde conjecture that the modular transformation corresponding to $\tau \to -1/\tau$ diagonalizes the fusion rules. If one can show that all irreducible highest weight representations at level $k$ of an affine Lie algebra indeed give a rational conformal field theory, the Verlinde formula which has attracted many mathematicians recently will be an easy consequence. But in fact the most difficult part is the rigorous construction of a rational conformal field theory satisfying the geometric axioms. Even in the affine Lie algebra case, a complete construction of the corresponding rational conformal field theories (not just the construction of the conformal blocks) is still to be given. On the other hand, the algebraic formulations of quantum field theories correspond to the operator product expansion approach in physics and many concrete quantum field theories in the algebraic for-mulations can be constructed rigorously from some simple algebraic data. But the axioms in the algebraic formulations are usually complicated and are conceptually difficult to understand. It is therefore necessary and important to establish rigorously the relationship between the algebraic and geometric formulations. In some sense, establishing the equivalence between the algebraic and geometric formulations can be thought of as establishing the existence of path integrals rigorously without writing down the classical fields and the Lagrangians of quantum field theories explicitly, since the axioms in the geometric formulations are nothing but the properties of path integrals. The papers [H1] [H2] [HL1] [HL2] [H4] are in this spirit. The present paper is another one in this spirit.

The present paper is based on the operadic formulation of (graded) vertex algebras in terms of the partial operad $\hat{K}$. Roughly speaking, a weak topological vertex algebra is a graded vertex algebra together with three operators $Q$, $g(0)$ and $g(-1)$ satisfying certain natural axioms including $Q^2 = 0$. A topological vertex algebra is a weak topological vertex algebra satisfying the additional axiom that $(g(0))^2$ is $Q$-exact. A strong topological vertex algebra is a topological vertex algebra satisfying the stronger axiom that $(g(0))^2 = 0$. Since a (weak, strong) topological vertex algebra is a graded vertex algebra, we have an action of the partial operad $\hat{K}$. Here we give a heuristic interpretation of two most important axioms on $Q$, $g(0)$ and $g(-1)$ in the case of strong topological vertex algebras: $[Q, g(l)] = L(l)$, $l = 0, -1$, where $Q$, $g(0)$ and $g(-1)$ are operators mentioned above and $L(0)$ and $L(-1)$ are operators defined using the underlying graded vertex algebra structure (see (2.9) and (2.10)). The operators $L(0)$ and $L(-1)$ can be thought of as linearly representing the Lie derivatives $\mathcal{L}_0$ and $\mathcal{L}_{-1}$ on tensor fields on $\hat{K}$ along certain directions. On $\hat{K}$, we have $[d, i_l] = \mathcal{L}_l$, $l = 0, -1$, where $d$ is the exterior derivation, $i_l$ is the interior derivation along the direction corresponing to $L(l)$. In general, for a graded vertex algebra, the algebraic structure on the space of differential forms does not have to be linearly represented. If we require that the algebraic structure on the space of differential forms is also linearly represented and assume that $d$, $i_0$ and $i_{-1}$ are represented by $Q$, $g(0)$ and $g(-1)$, respectively, then we must have $[Q, g(l)] = L(l)$, $l = 0, -1$, which are exactly the two axioms we would like to interpret.

The main technical work in the present paper is the proofs of the properties of the forms $\omega_j$, $j \in \mathbb{N}$, in Section 3. Our construction of $\omega_j$ (see (3.21) and (3.20)) is motivated by [Z]. In fact in the case that the (strong) topological vertex algebra is a weak vertex operator algebra of central charge 26 tensored with the graded vertex operator algebra of the ghosts, the forms $\omega_j$ in the present paper are completely analogous to the forms in [Z]. The equivalence proved in Section 4 between the notion of (weak, strong) topological vertex algebras and the operadic formulation in terms of these forms in fact shows that any family of forms satisfies those properties in Section 3 must be of the form (3.21) and (3.20). In the case of strong topological vertex algebras the axiom $(g(0))^2 = 0$ and its consequences $(g(-1))^2 = g(0)g(-1) + g(-1)g(0) = 0$ simplify the definition of the forms and consequently simplify the proofs greatly. In this paper, for completeness and for the possible use in the future, we are interested in arbitrary topological vertex algebras and also in weak topological vertex algebras. Thus we cannot assume these relations. Consequently the proofs for (weak) topological vertex algebras are much more complicated than those for strong topological vertex algebras.

This paper is organized as follows: In Section 2 we beriefly discuss the notions of graded vertex operator algebra, weak (graded) vertex operator algebra, topological vertex operator algebra, (graded) vertex algebra, and introduce the notions of weak topological vertex algebra, topological vertex algebra and strong topological vertex algebra. We also discuss the action of the partial operad $\hat{K}$ on (graded) vertex algebras. In the subsequent section we construct the forms $\omega_j$ from the data of a (weak, strong) topological vertex algebra and prove their basic properties. We use these forms and their properties to define the notions of (weak, strong) topological $\hat{K}$-associative algebra and show that these notions are equivalent to those of (weak, strong) topological vertex algebra in Section 4. In the same section, we also show that there is a family of natural symmetry-group-equivariant chain maps from the differential graded partial operad of the differentiable chain complex of $\hat{K}$ to the differential graded endomorphism partial pseudo-operad of a weak topological vertex algebra and that a strong topological vertex algebra is an algebra over the differential graded partial operad of the differentiable chain complex of $\hat{K}$. In Section 5, we first review the definition of Batalin-Vilkovisky algebra and Gerstenhaber algebra, some basic results including Cohen's and Getzler's theorem and some other related notions and results. Then combining the results obtained in Section 3 and in Section 4 with Getzler's theorem (or Cohen's theorem), we obtain the Batalin-Vilkovisky algebra structure (or the Gerstenhaber algebra structure) on the cohomology of a topological vertex algebra (or a weak topological vertex algebra).

**Acknowledgement.** It is a pleasure to express my gratitude to Jim Stasheff for many illuminating comments and discussions. I would also like to thank Murray Gerstenhaber, James Lepowsky, Bong H. Lian, Takashi Kimura, Alexander Voronov and Gregg Zuckerman for useful discussions.

## 2. Topological vertex algebras and the action of a (partial) operad

In this section, we discuss various variants of vertex operator algebras, introduce the notions of (weak, strong) topological vertex algebra and describe the actions of a (partial) operad on these variants of vertex operator algebras. We will only give definitions and state the theorems. All proofs are omitted since they are either easy exercise or to be published elsewhere.

We begin with the definition of graded vertex operator algebra (or "super chiral algebra" as is called in [LZ]). The notion of graded vertex operator algebra is a natural generalization of that of vertex operator algebra and can be viewed as a specialization of the notion of abelian intertwining algebra introduced in [DL]. We assume that the readers are familiar with the definition of vertex operator algebra, as presented in [FLM2] or [FHL]. We only give the differences between vertex operator algebras and graded vertex operator algebras.

For a *graded vertex operator algebra*, the vector space $V$ is $\mathbb{Z} \times \mathbb{Z}$-graded (graded by weights and by *fermion numbers*), that is,

$$V = \coprod_{m,n\in\mathbb{Z}} V_{(n)}^{(m)} = \coprod_{n\in\mathbb{Z}} V_{(n)} = \coprod_{m\in\mathbb{Z}} V^{(m)} \tag{2.1}$$

where

$$V_{(n)} = \coprod_{m\in\mathbb{Z}} V_{(n)}^{(m)}, \quad V^{(m)} = \coprod_{n\in\mathbb{Z}} V_{(n)}^{(m)}. \tag{2.2}$$

The map $Y : V \otimes V \to V[[x, x^{-1}]]$ maps $V^{(m_1)} \otimes V^{(m_2)}$ to $V^{(m_1+m_2)}[[x, x^{-1}]]$. For $v \in V^{(m)}$, we say that $v$ has fermion number $m$ and use $|v|$ to denote $m$. For the vacuum $\mathbf{1}$ and the Virasoro element $\omega$, $|\mathbf{1}| = |\omega| = 0$. The (Cauchy-)Jacobi identity is of the following form for $u, v$ with homogeneous fermion numbers:

$$\begin{aligned} x_0^{-1}\delta\left(\frac{x_1 - x_2}{x_0}\right) Y(u, x_1)Y(v, x_2) - (-1)^{|u||v|}x_0^{-1}\delta\left(\frac{x_2 - x_1}{-x_0}\right) Y(v, x_2)Y(u, x_1) \\ = x_2^{-1}\delta\left(\frac{x_1 - x_0}{x_2}\right) Y(Y(u, x_0)v, x_2). \end{aligned} \tag{2.3}$$

All the other data and axioms are the same as those for vertex operator algebras. We also call a quadruple $(V, Y, \mathbf{1}, \omega)$ satisfying all axioms for vertex operator algebras (or for graded vertex operator algebras) except the two grading axioms — $\dim V_n < \infty$ and $V_{(n)} = 0$ for $n$ sufficiently small — a *weak vertex operator algebra* (or a *weak graded vertex operator algebra*).

Next we recall the definition of topological vertex operator algebra (or "topological chiral algebra") given in [LZ]. A *topological vertex operator algebra* is a weak graded vertex operator algebra $V$ together with three distinguished elements $f \in V_{(1)}$, $q \in V_{(1)}^{(1)}$ and $g \in V_{(2)}^{(1)}$ satisfying the following axioms:

(i) Let $Y(f, x) = \sum_{n\in\mathbb{Z}} f_n x^{-n-1}$. Then for any $v \in V^{(m)}$,

$$f_0 v = mv. \tag{2.4}$$

(ii) Let $Y(q, x) = \sum_{n\in\mathbb{Z}} q_n x^{-n-1}$ and $Q = q_0$. Then

$$L(n)q = 0, \quad n > 0, \tag{2.5}$$

$$Q^2 = 0. \tag{2.6}$$

(iii) Let $Y(g, x) = \sum_{n\in\mathbb{Z}} g(n)x^{-n-2}$ and $\omega$ the Virasoro element of $V$. Then

$$L(n)g = 0, \quad n > 0, \tag{2.7}$$

$$Qg = \omega. \tag{2.8}$$

Note that (2.4) implies $f \in V^{(0)}_{(1)}$. In fact, the construction of the Batalin-Vilkovisky algebra (or the "coboundary Gerstenhaber algebra") structure on the cohomology of a topological vertex operator algebra in [LZ] also uses the additional axiom that $(g(0))^2$ is $Q$-exact, though the construction of the Gerstenhaber algebra structure does not need this axiom.

Examples of topological operator vertex algebras are weak vertex operator algebras with cenral charge 26 tensored with the graded vertex operator algebra of ghosts and the holomorphic genus-zero parts of $N = 2$ twisted superconformal theories. These examples all satisfy the additional axioms $(g(0))^2 = (g(-1))^2 = g(0)g(-1) + g(-1)g(0) = 0$.

In the construction of the Gerstenhaber algebra or the Batalin-Vilkovisky algebra structure on the cohomology of a topological vertex operator algebra by Lian and Zuckerman [LZ], only a small part of the axioms for topological vertex operator algebras is needed. In this paper we also need only a small part of the axioms. We summarize those axioms which we do need in the definition of the notions of (weak, strong) topological vertex algebra below.

Before giving the definition, we need to discuss certain variants of vertex operator algebras and graded vertex operator algebras: the notions of vertex algebra and of graded vertex algebra. The notion of vertex algebra is introduced by Borcherds in [B]. In [L] they are called preVOAs. (In [DL], vertex algebra is a technical term referred to weak vertex operator algebras mentioned above.) In this paper, by a *vertex algebra* we mean a $\mathbb{Z}$-graded vector apace $V$ equipped with a map $Y : V \otimes V \to V[[x, x^{-1}]]$, $u \otimes v \mapsto Y(u, x)v = \sum_{n\in\mathbb{Z}} u_n v x^{-n-1}$ and a *vacuum* $\mathbf{1}$ such that for any homogeneous element $v$, $v_k$ maps $V_{(n)}$ to $V_{(n+\mathrm{wt}\, v-k-1)}$, satisfying all the axioms for a vertex operator algebra except the axioms $\dim V_{(n)} < \infty$, $V_{(n)} = 0$ for $n$ sufficiently small and those axioms for the Virasoro element. The definition of *graded vertex algebra* is obvious. Given a vertex algebra or a graded vertex algebra, we can define two operators $L(0)$ and $L(-1)$ as follows: For any $v \in V_{(n)}$,

$$L(0)v = nv, \tag{2.9}$$

and for any $v \in V$,

$$L(-1)v = \lim_{x\to 0} \frac{d}{dx} Y(v, x)\mathbf{1}. \tag{2.10}$$

Using these definitions and the (Cauchy-)Jacobi identity, it is easy to verify the

following identities:

$$[L(0), L(-1)] = L(-1), \tag{2.11}$$

$$[L(-1), Y(v, x)] = Y(L(-1)v, x) = \frac{d}{dx}Y(v, x), \tag{2.12}$$

$$[L(0), Y(v, x)] = Y((L(0) + xL(-1))v, x) = Y(L(0)v, x) + x\frac{d}{dx}Y(v, x). \tag{2.13}$$

**Definition 2.1.** A *weak topological vertex algebra* is a graded vertex algebra together with operators $Q$, $g(0)$ and $g(-1)$ mapping $V^{(m)}$ to $V^{(m+1)}$ $V^{(m-1)}$ and $V^{(m-1)}$, respectively, satisfying all the axioms for topological vertex operator algebras which still make sense, that is, satisfying

$$Q^2 = 0, \tag{2.14}$$
$$[Q, g(0)] = L(0), \tag{2.15}$$
$$[Q, g(-1)] = L(-1) \tag{2.16}$$
$$[L(0), g(0)] = 0, \tag{2.17}$$
$$[L(0), g(-1)] = g(-1), \tag{2.18}$$
$$[L(-1), g(0)] = -g(-1), \tag{2.19}$$
$$[L(-1), g(-1)] = 0, \tag{2.20}$$
$$QY(u, x) - (-1)^{|u|}Y(u, x)Q = Y(Qu, x), \tag{2.21}$$
$$g(0)Y(u, x) - (-1)^{|u|}Y(u, x)g(0) = Y((g(0) + xg(-1))u, x), \tag{2.22}$$
$$g(-1)Y(u, x) - (-1)^{|u|}Y(u, x)g(-1) = Y(g(-1)u, x) \tag{2.23}$$

where the bracket $[A.B]$ of two operators $A$ and $B$ with homogeneous fermion numbers is defined by

$$[A, B] = AB - (-1)^{|A||B|}BA. \tag{2.24}$$

A *topological vertex algebra* is a weak topological vertex algebra such that $(g(0))^2$ is a $Q$-exact operator, that is, there exists an operator $U_{0,0} : V \to V$ such that

$$(g(0))^2 = [Q, U_{0,0}]. \tag{2.25}$$

A *strong topological vertex algebra* is a weak topological vertex algebra satisfying

$$(g(0))^2 = 0. \tag{2.26}$$

From (2.15) — (2.20) and the Jacobi identity for the bracket $[\cdot,\cdot]$ we see that for a weak topological vertex algebra,

$$(g(-1))^2 = \frac{[Q,[g(0),(g(-1))^2]]}{2} = -\frac{[Q,[g(-1),[g(0),g(-1)]]]}{2}, \tag{2.27}$$

$$[g(0),g(-1)] = [Q,[g(0),[g(0),g(-1)]]] = -[Q,[g(-1),(g(0))^2]]. \tag{2.28}$$

In particular, these formulas also hold for a topological vertex algebra. From these formulas we see that for a strong topological vertex algebra we also have

$$[g(0),g(-1)] = 0 \tag{2.29}$$

and

$$(g(-1))^2 = 0. \tag{2.30}$$

It is also easy to show that a topological vertex operator algebra is a weak topological vertex algebra using the (Cauchy-)Jacobi identity. When there is no confusion, we will call $V$ a (weak, strong) topological vertex algebra.

Now we discuss the action of a (partial) operad on graded vertex operator algebras and in particular on topological vertex operator algebras. For the basic notions in the language of (partial) operads, see [May], [HL1] or [HL2]. Let $K(j)$, $j\in\mathbb{N}$, be the moduli space of one-dimensional genus-zero compact connected complex manifolds with $j+1$ ordered punctures, the zeroth negatively oriented and the others positively oriented, and with local analytic coordinates vanishing at these punctures. The family $K=\{K(j)|j\in\mathbb{N}\}$ is an associative analytic $\mathbb{C}^\times$-rescalable partial operad [HL1] [HL2] [H4]. Using certain one-dimensional extensions of this partial operad, the notion of vertex associative algebra with central charge $c$ is introduced [HL1] [HL2]. It has been established [H1] [H2] [HL1] [HL2] [H4] that the category of vertex operator algebras with central charge $c$ is isomorphic to the category of vertex associative algebras with central charge $c$. In particular, for any vertex operator algebra $V$, there is a projective action of the partial operad $K$. See [HL1] and [HL2] for more detailed descriptions.

These discussions can be generalized to graded vertex operator algebras without any difficulty. Given a $\mathbb{Z}\times\mathbb{Z}$-graded vector space $V$ (graded by weights and fermion numbers) and a $\mathbb{Z}\times\mathbb{Z}$-graded subspace $W$, we can define the graded endomorphism partial pseudo-operad $\mathcal{H}_{V,W}$ in the same way as in [HL1] or [HL2] except that the left actions of the symmetry groups are defined such that for any $f\in\mathcal{H}_{V,W}(j)$ and $\sigma_{i,i+1}\in S_j$ which is the transposition permuting $i$ and $i+1$,

$$(\sigma_{i,i+1}(f))(v_1,\dots,v_i,v_{i+1},\dots,v_j) = (-1)^{|v_i||v_{i+1}|}f(v_1,\dots,v_{i+1},v_i,\dots,v_j) \tag{2.31}$$

for any $v_1,\dots,v_j\in V$ with homogeneous fermion numbers. Using this graded endomorphism partial pseudo-operad, we can define graded vertex associative algebra in the same way as that defining vertex associative algebras. Then the proof in the case of vertex operator algebras can be adopted to show that the category of graded vertex operator algebras with central charge $c$ is isomorphic to the category of graded vertex associative algebras with central charge $c$. In particular, given any graded vertex operator algebra, there is a projective action of the partial operad $K$.

The partial operad $K$ has several important partial suboperads. The two which are useful in the present paper are $\hat{K} = \{\hat{K}(j) | j \in \mathbb{N}\}$ where $\hat{K}(j)$ consists of those conformal equivalence classes whose members are conformally equivalent to $\mathbb{C} \cup \{\infty\}$ with the negatively oriented puncture $\infty$, the positively oriented punctures $z_1, \dots, z_j$, and with the standard local coordinates at $\infty$, the standard local coordinates multiplied by nonzero complex numbers as local coordinates at $z_1, \dots, z_j$, and $\overline{K}$ which is the same as $\hat{K}$ except that the local coordinates at $z_1, \dots, z_j$ are also standard. The family $\hat{K}$ is also an associative analytic $\mathbb{C}$-rescalable partial operad and $\overline{K}$ is a partial suboperad of $\hat{K}$. For any graded vertex algebra $V$, there is an action of the partial operad $\hat{K}$. In fact this action can be written down easily. Let $P \in \hat{K}(j)$. Then $P$ can be identified with $(z_1, \dots, z_j; a_1, \dots, a_j) \in \mathbb{F}_j(\mathbb{C}) \times (\mathbb{C}^\times)^j$, where $\mathbb{F}_j(\mathbb{C}) = \{(z_1, \dots, z_j) \in (\mathbb{C}^\times)^j | z_k \neq z_l,\ 1 \leq k < l \leq j\}$. For convenience, we call $z_1, \dots, z_j$ the *punctures* of $P$ and $a_1, \dots, a_j$ the *local coordinates* of $P$. Let $V$ be a graded vertex algebra and $W = 0$. In this case $\mathcal{H}_{V,0}(j) = \mathrm{Hom}(V^{\otimes j}, \overline{V})$ where $\overline{V} = \prod_{n \in \mathbb{Z}} V_{(n)}$. We define $\nu(P) \in \mathcal{H}_{V,0}$ by

$$\nu(P)(v_1 \otimes \cdots \otimes v_j) = Y((a_1)^{-L(0)}v_1, x_1) \cdots Y((a_j)^{-L(0)}v_j, x_j)\mathbf{1}\big|_{x_1 = z_1, \dots, x_j = z_j} \tag{2.32}$$

when $|z_1| > \cdots > |z_j|$. When this inequality does not hold, we can change the order of the vertex operators or use the iterations of the vertex operators or use analytic extension to define $\nu(P)$. We have

**Proposition 2.1.** *The maps $\nu : \hat{K}(j) \to \mathcal{H}_{V,0}(j)$, $j \in \mathbb{N}$ give a morphism of partial pseudo-operad such that the image is a partial operad. In particular, the maps $\nu$ give a morphism of partial operads from $\hat{K}$ to its image.*

The proof of this proposition uses only (2.13) and the duality properties of graded vertex algebras.

**Definition 2.2.** A *graded meromorphic $\hat{K}$-associative algebra* is a $\mathbb{Z} \times \mathbb{Z}$-graded vector space $V$ of the form (2.1) together with a morphism $\nu$ of partial pseudo-operads from $\hat{K}$ to $\mathcal{H}_{V,0}$ such that for any $v \in V_{(n)}$

$$\nu((0; a))v = a^{-n}v \tag{2.33}$$

(where $(0; a) \in \mathbb{F}_1(\mathbb{C}) \times \mathbb{C}^\times = \hat{K}(1)$) and such that for any $v_1, \dots, v_j \in V$, $v' \in V' = \coprod_{n \in \mathbb{Z}} V^*_{(n)}$ and for $P \in \hat{K}(j) = \mathbb{F}_j(\mathbb{C}) \times (\mathbb{C}^\times)^j$ of the form $(z_1, \dots, z_j; 1, \dots, 1)$, $\langle v', \nu(P)(v_1, \dots, v_j)\rangle$ as a function of $z_1, \dots, z_j$ is meromorphic on $\mathbb{F}_j(\mathbb{C})$ with $z_i = z_k$, $i < k$, and $z_i = \infty$ as the only possible poles, and for fixed $v_i, v_k \in V$ there is an upper bound for the orders of the pole $z_i = z_k$ of the functions $\langle v', \nu(P)(v_1, \dots, v_j)\rangle$ for all $v_1, \dots, v_{i-1}, v_{i+1}, \dots, v_{k-1}, v_{k+1}, \dots, v_j \in V$, $v' \in V'$.

Part of the proof of the equivalence theorem for graded vertex operator algebras in fact proves the following equivalence theorem for graded vertex algebras:

**Theorem 2.2.** *The functor given by $(V, Y, \mathbf{1}) \to (V, 0, \nu)$ is an isomorphism from the category of graded vertex algebras to the category of graded meromorphic $\hat{K}$-associative algebras.*

## 3. The meromorphic forms constructed from a (weak, strong) topological vertex algebras

Now we would like to see what kind of topological and geometric information the extra data and axioms in the definitions of (strong, weak) topological vertex algebra give. Let $V$ be a weak topological vertex algebra. We define a holomorphic form $\omega_j \in \Omega^*(\hat{K}(j), \mathcal{H}_{V,0}(j))$ valued in the space $\mathcal{H}_{V,0}(j)$ for each $j \in \mathbb{N}$. Since holomorphic forms valued at a point are multilinear skew-symmetric maps from products of the holomorphic tangent space at this point to $\mathbb{C}$, we first have to discuss the holomorphic tangent space of $\hat{K}(j)$. Recall that $\hat{K}(j)$ can be identified with $\mathbb{F}_j(\mathbb{C}) \times (\mathbb{C}^\times)^j$ whose elements are denoted $(z_1, \dots, z_j; a_1, \dots, a_j)$. Therefore a holomorphic tangent vector $X_P$ at the point $P = (z_1, \dots, z_j; a_1, \dots, a_j) \in \hat{K}(j)$ can be written in the form

$$X_P = c_{-1}^{(1)}(-a_1^{-1}\frac{\partial}{\partial z_1}\big|_P) + \cdots + c_{-1}^{(j)}(-a_j^{-1}\frac{\partial}{\partial z_j}\big|_P) + c_0^{(1)}(-a_1\frac{\partial}{\partial a_1}\big|_P) + \cdots + c_0^{(j)}(-a_j\frac{\partial}{\partial a_j}\big|_P). \tag{3.1}$$

where $c_{-1}^{(1)}, \dots, c_{-1}^{(j)}, c_0^{(1)}, \dots, c_0^{(j)} \in \mathbb{C}$. Let

$$X_P^{(i)} = c_{-1}^{(i)}(-a_i^{-1}\frac{\partial}{\partial z_i}\big|_P) + c_0^{(i)}(-a_i\frac{\partial}{\partial a_i}\big|_P), \quad 1 \le i \le j. \tag{3.2}$$

Then

$$X_P = \sum_{i=1}^{j} X_P^{(i)}. \tag{3.3}$$

We denote the space of holomorphic tangent vectors at $P \in \hat{K}(j)$ by $T_P(\hat{K}(j))$, the subspace of $T_P(\hat{K}(j))$ spanned by $-a_i^{-1}\frac{\partial}{\partial z_i}\big|_P$ and $-a_i\frac{\partial}{\partial a_i}\big|_P$ by $T_P^{(i)}(\hat{K}(j))$, $i = 1, \cdots, j$. Then $\dim_{\mathbb{C}} T_P(\hat{K}(j)) = 2j$, $\dim_{\mathbb{C}} T_P^{(i)}(\hat{K}(j)) = 2$, $i = 1, \dots, j$, where $\dim_{\mathbb{C}}$ means the complex dimension. And we have

$$T_P(\hat{K}(j)) = \bigoplus_{i=1}^{j} T_P^{(i)}(\hat{K}(j)). \tag{3.4}$$

Let $T(\hat{K}(j))$ be the holomorphic bundle over $\hat{K}(j)$ whose fiber at $P \in \hat{K}(j)$ is $T_P(\hat{K}(j))$ and $T_P^{(i)}(\hat{K}(j))$, $i = 1, \dots, j$, the holomorphic bundles over $\hat{K}(j)$ whose fibers at $P \in \hat{K}(j)$ are $T_P^{(i)}(\hat{K}(j))$. Then we also have a decomposition

$$T(\hat{K}(j)) = \bigoplus_{i=1}^{j} T^{(i)}(\hat{K}(j)). \tag{3.5}$$

Any holomorphic section $X^{(i)}$ of $T^{(i)}(\hat{K}(j))$ for $i = 1, \dots, j$ is of the form

$$X^{(i)} = c_{-1}^{(i)}(-a_i^{-1}\frac{\partial}{\partial z_i}) + c_0^{(i)}(-a_i\frac{\partial}{\partial a_i}) \tag{3.6}$$

where $c_{-1}^{(i)}$ and $c_0^{(i)}$ are analytic functions on $\hat{K}(j)$. For any holomorphic section $X$ of $T(\hat{K}(j))$, from (3.5) we have a decomposition

$$X = \sum_{i=1}^{j} X^{(i)} \tag{3.7}$$

where $X^{(i)}$, $i = 1, \dots, j$, are holomorphic sections of $T^{(i)}(\hat{K}(j))$, $i = 1, \dots, j$, respectively. We denote the spaces of holomorphic sections of $T^{(i)}(\hat{K}(j))$, $i = 1, \dots, j$, and $T(\hat{K}(j))$ by $\Gamma(T^{(i)}(\hat{K}(j)))$, $i = 1, \cdots, j$, and by $\Gamma(T(\hat{K}(j)))$, respectively. Then by (3.7)

$$\Gamma(T(\hat{K}(j))) = \bigoplus_{i=1}^{j} \Gamma(T^{(i)}(\hat{K}(j))). \tag{3.8}$$

We can discuss the holomorphic tangent vector space $T_P(\overline{K}(j))$ at $P \in \overline{K}(j)$, the holomorphic tangent bundle $T(\overline{K}(j))$ and the space $\Gamma(T(\overline{K}(j)))$ of holomorphic sections of $T(\overline{K}(j))$ for each $j \in \mathbb{N}$ in the same way. In particular, we also have $T_P^{(i)}(\overline{K}(j))$ and $T^{(i)}(\overline{K}(j))$, $1 \le i \le j$, $j \in \mathbb{N}$ defined in the obvious way. All these vector spaces and vector bundles can be embedded naturally in the corresponding vector spaces and vector bundles for $\hat{K}(j)$.

We define $4j$ operators $g^{(i)}(0)$, $g^{(i)}(-1)$, $L^{(i)}(0)$ and $L^{(i)}(-1)$, $i = 1, \dots, j$, on $V^{\otimes j}$ by

$$g^{(i)}(0)(v_1 \otimes \cdots \otimes v_i \otimes \cdots \otimes v_j) = (-1)^{|v_1|+\cdots|v_{i-1}|} v_1 \otimes \cdots \otimes g(0)v_i \otimes \cdots \otimes v_j, \tag{3.9}$$

$$g^{(i)}(-1)(v_1 \otimes \cdots \otimes v_i \otimes \cdots \otimes v_j) = (-1)^{|v_1|+\cdots|v_{i-1}|} v_1 \otimes \cdots \otimes g(-1)v_i \otimes \cdots \otimes v_j \tag{3.10}$$

$$L^{(i)}(0)(v_1 \otimes \cdots \otimes v_j) = v_1 \otimes \cdots \otimes L(0)v_i \otimes \cdots \otimes v_j, \tag{3.11}$$

$$L^{(i)}(-1)(v_1 \otimes \cdots \otimes v_j) = v_1 \otimes \cdots \otimes L(-1)v_i \otimes \cdots \otimes v_j. \tag{3.12}$$

From these definitions and the formulas (2.11), (2.17) — (2.20), we see that they satisfy

$$[g^{(k)}(p), g^{(l)}(q)] = g^{(k)}(p)g^{(l)}(q) + g^{(l)}(q)g^{(k)}(p) = 0, \quad p, q = 0, -1, \quad k \neq l, \tag{3.13}$$

$$\begin{aligned}[L^{(k)}(p), L^{(l)}(q)] &= L^{(k)}(p)L^{(l)}(q) - L^{(l)}(q)L^{(k)}(p) \\ &= \delta_{kl}(p-q)L^{(k)}(p+q), \qquad\qquad p, q = 0, -1,\end{aligned} \tag{3.14}$$

$$\begin{aligned}[L^{(k)}(p), g^{(l)}(q)] &= L^{(k)}(p)g^{(l)}(q) - g^{(l)}(q)L^{(k)}(p) \\ &= \delta_{kl}(p-q)L^{(k)}(p+q), \qquad\qquad p, q = 0, -1.\end{aligned} \tag{3.15}$$

For a fixed $i$, $1 \leq i \leq j$, let $X_P^{(i)} \in T_P^{(i)}$. We define the operators $g(X_P^{(i)})$ on $V^{\otimes j}$ by replacing $-a_i \frac{\partial}{\partial a_i}\big|_P$ and $-a_i^{-1}\frac{\partial}{\partial z_i}\big|_P$ in $X_P^{(i)}$ by $g^{(i)}(0)$ and $g^{(i)}(-1)$, respectively, and define the operators $L(X_P^{(i)})$ on $V^{\otimes j}$ by replacing $-a_i \frac{\partial}{\partial a_i}\big|_P$ and $-a_i^{-1}\frac{\partial}{\partial z_i}\big|_P$ in $X_P^{(i)}$ by $L^{(i)}(0)$ and $L^{(i)}(-1)$, respectively. For $X_P \in T_P(\hat{K}(j))$ of the form (3.3), we define

$$g(X_P) = \sum_{i=1}^{j} g(X_P^{(i)}) \tag{3.16}$$

$$L(X_P) = \sum_{i=1}^{j} L(X_P^{(i)}). \tag{3.17}$$

For holomorphic sections $X^{(i)}$ of $T^{(i)}(\hat{K}(j))$ and $X$ of $T(\hat{K}(j))$, we can define $g(X^{(i)})$, $g(X)$, $L(X^{(i)})$ and $L(X)$ in the obvious way. Using (2.12) and the definition of $\nu(P)$, we have

$$X\nu(P) = \nu(P)L(X). \tag{3.18}$$

Given $X_{1,P}, \ldots, X_{n,P} \in T_P(\hat{K}(j))$, we use the following notation:

$$g(X_{1,P}) \wedge \cdots \wedge g(X_{n,P}) = \frac{1}{n!}\sum_{\sigma \in S_n} (-1)^{\mathrm{sgn}\sigma} g(X_{\sigma(1),P}) \cdots g(X_{\sigma(n),P}). \tag{3.19}$$

We are ready to define the form $\omega_j$. For any nonnegative integer $n$, we define a holomorphic form $\omega_{j,n}$ of degree $n$ by

$$\omega_{j,n}\big|_P(X_{1,P}, \ldots, X_{n,P}) = \nu(P) g(X_{1,P}) \wedge \cdots \wedge g(X_{n,P}) \tag{3.20}$$

where $P \in \hat{K}$ and $X_{1,P}, \ldots, X_{n,P}$ are tangent vectors at $P$. From the definition, $\omega_{j,n}$ is skew-symmetric and is holomorphic in $P$. Thus it indeed gives a holomorphic form.

It is obvious that $\omega_{j,n} = 0$ for $n > 2j = \dim_{\mathbb{C}} \hat{K}(j)$. The holomorphic form $\omega_j$ is defined to be the sum of $\omega_{j,n}$ for all $n$, that is,

$$\omega_j = \sum_{n=0}^{2j} \omega_{j,n}. \tag{3.21}$$

From the definition and (2.33) we have

$$\omega_{1,0}\big|_{(0,a)} v = a^{-n} v \tag{3.22}$$

for any $v \in V_{(n)}$.

The differential form $\omega_j$ has four important properties. The first is its meromorphicity.

**Proposition 3.1.** *For any integers $m$, $i_1, \dots, i_m$, $j_1, \dots, j_{n-m}$ satisfying $1 \le m \le n$, $1 \le i_1 < \cdots < i_m \le j$, $1 \le j_1 < \cdots < j_{n-m} \le j$, and for any $v_1, \dots, v_j \in V$, $v' \in V'$,*

(3.23)

$$\langle v', \omega_{j,n}(-a_{i_1}^{-1}\frac{\partial}{\partial z_{i_1}}, \cdots, -a_{i_1}^{-1}\frac{\partial}{\partial z_{i_m}}, -a_{j_1}\frac{\partial}{\partial a_{j_1}}, \cdots, -a_{j_{n-m}}\frac{\partial}{\partial a_{j_{n-m}}})(v_1 \otimes \cdots \otimes v_j)\rangle$$

*as a function of $P$ is meromorphic on $\mathbb{F}_j(\mathbb{C})$ with $z_i = z_k$, $i < k$, and $z_i = \infty$ as the only possible poles, and for fixed $v_i, v_k \in V$ there is an upper bound for the orders of the pole $z_i = z_k$ of the functions (3.23) for all $v_1, \dots, v_{i-1}$, $v_{i+1}, \dots, v_{k-1}$, $v_{k+1}, \dots, v_j \in V$, $v' \in V'$.*

*Proof.* This proposition follows directly from the the definition of $\omega_{j,n}$ , Definition 2.2 and Theorem 2.2. □

The second property concerns the exterior derivatives of these meromorphic forms. Before stating the second property, we have to define the action of the operator $Q$ on the space $\mathcal{H}_{V,0}(j)$. The fermion number grading on $V$ induces $\mathbb{Z}$-gradings on $V^{\otimes j}$ and $\overline{V}$. These $\mathbb{Z}$-gradings induce a $\mathbb{Z}$-grading on $\mathrm{Hom}(V^{\otimes j}, \overline{V}) = \mathcal{H}_{V,0}(j)$. We still call these gradings the *fermion number gradings.* The operator $Q$ on $V$ extends to an operator (still denoted $Q$) on $\overline{V}$. For any $v_1, \dots, v_j \in V$ with homogeneous fermion numbers, we define

$$Q_{V^{\otimes j}}(v_1 \otimes \cdots \otimes v_j) = \sum_{i=1}^{j} (-1)^{|v_1| + \cdots + |v_{i-1}|} h(v_1 \otimes \cdots \otimes Qv_i \otimes \cdots \otimes v_j). \tag{3.24}$$

Using the linearity we extend $Q_{V^{\otimes j}}$ to an operator on $V^{\otimes j}$. Let $h \in \mathcal{H}_{V,0}(j) = \mathrm{Hom}(V^{\otimes j}, \overline{V})$ with fermion number $n$. We define $Q_{\mathcal{H}_{V,0}} h \in \mathcal{H}_{V,0}(j) = \mathrm{Hom}(V^{\otimes j}, \overline{V})$ by

(3.25)

$$(Q_{\mathcal{H}_{V,0}}(h))(v_1 \otimes \cdots \otimes v_j) = Q(h(v_1 \otimes \cdots \otimes v_j)) - (-1)^n h(Q_{V^{\otimes}}(v_1 \otimes \cdots \otimes v_j)).$$

Using (2.21), the formula

$$[Q, L(0)] = 0 \tag{3.26}$$

(a consequence of (2.15)) and the definition of $\nu$, we have

$$Q\nu(P) = \nu(P)Q_{V^{\otimes j}}. \tag{3.27}$$

From the condition that $Y$ maps $V^{(m_1)} \otimes V^{(m_2)}$ to $V^{(m_1+m_2)}[[x, x^{-1}]]$ and the definition of $\nu$, we see that the fermion number of $\nu(P)$ for any $P \in \hat{K}$ is 0. Thus (3.27) is equivalent to

$$Q_{\mathcal{H}_{V,0}}(\nu(P)) = 0. \tag{3.28}$$

**Proposition 3.2.** *The differential form $\omega_j$ satisfies the following equation:*

$$d\omega_j = Q_{\mathcal{H}_{V,0}}(\omega_j). \tag{3.29}$$

*Proof.* The formula (3.29) is equivalent to

$$d\omega_{j,n} = Q_{\mathcal{H}_{V,0}}(\omega_{j,n+1}) \tag{3.30}$$

for any $n \geq 0$. Since when evaluated at $P \in \hat{K}(j)$, both sides of (3.30) are multilinear skew-symmetric maps from $(T_P(\hat{K}(j)))^n$ to $\mathbb{C}$, we need only to show

$$d\omega_{j,n}(X_1^{(i_1)}, \dots, X_{n+1}^{(i_{n+1})}) = Q_{\mathcal{H}_{V,0}}(\omega_{j,n+1}(X_1^{(i_1)}, \dots, X_{n+1}^{(i_{n+1})})) \tag{3.31}$$

for $X_k^{(i_k)} \in \Gamma(T^{(i_k)}(\hat{K}(j)))$, $k = 1, \dots, n+1$, $1 \leq i_1 \leq \cdots \leq i_{n+1} \leq j$ equal to either $-a_{i_k}^{-1}\frac{\partial}{\partial z_{i_k}}$ or $-a_{i_k}\frac{\partial}{\partial a_{i_k}}$. Since $\dim_{\mathbb{C}} T_P^{(i_k)}(\hat{K}(j)) = 2$, $k = 1, \dots, n+1$, both sides of (3.31) are zero if three of $i_1, \dots, i_{n+1}$ are the same. Thus we can assume that there are at least two different numbers in any three of $i_1, \dots, i_{n+1}$. Using the definition of the exterior derivative and the definition of $\omega_{j,n}$, we have

(3.32)
$$
\begin{aligned}
&(d\omega_{j,n})(X_1^{(i_1)}, \dots, X_{n+1}^{(i_{n+1})})\\
&\quad= \sum_{k=1}^{n+1}(-1)^{k+1}X_k^{(i_k)}\omega_{j,n}(X_1^{(i_1)}, \dots, X_{k-1}^{(i_{k-1})}, X_{k+1}^{(i_{k+1})}, \dots, X_{n+1}^{(i_{n+1})})\\
&\quad\quad+ \sum_{1\le k<l\le n+1}(-1)^{k+l}\omega_{j,n}([X_k^{(i_k)}, X_l^{(i_l)}], X_1^{(i_1)}, \dots, X_{k-1}^{(i_{k-1})}, X_{k+1}^{(i_{k+1})},\\
&\quad\quad\quad\quad \dots, X_{l-1}^{(i_{l-1})}, X_{l+1}^{(i_{l+1})}, \dots, X_{n+1}^{(i_{n+1})})\\
&\quad= \sum_{k=1}^{n+1}(-1)^{k+1}X_k^{(i_k)}\nu(P)\frac{1}{n!}\sum_{\sigma\in S_{n+1}, \sigma(k)=k}(-1)^{\mathrm{sgn}\sigma}g(X_{\sigma(1)}^{(i_{\sigma(1)})})\cdots g(X_{\sigma(k-1)}^{(i_{\sigma(k-1)})})\cdot\\
&\quad\quad\quad\quad \cdot g(X_{\sigma(k+1)}^{(i_{\sigma(k+1)})})\cdots g(X_{\sigma(n+1)}^{(i_{\sigma(n+1)})})\\
&\quad\quad+ \sum_{1\le l<k\le n+1}(-1)^{l+k}\nu(P)\frac{1}{n!}\sum_{p\ne l,k}\sum_{\substack{\sigma\in S_{n+2}, \sigma(l+1)=l+1\\ \sigma(k+1)=k+1, \sigma^{-1}(1)=p}}(-1)^{\mathrm{sgn}\sigma}\cdot\\
&\quad\quad\quad \cdot g(X_{\sigma(1)-1}^{(i_{\sigma(1)-1})})\cdots g(X_{\sigma(p-1)-1}^{(i_{\sigma(p-1)-1})})g([X_l^{(i_l)}, X_k^{(i_k)}])g(X_{\sigma(p+1)-1}^{(i_{\sigma(p+1)-1})})\cdots g(X_{\sigma(l)-1}^{(i_{\sigma(l)-1})})\cdot\\
&\quad\quad\quad \cdot g(X_{\sigma(l+2)-1}^{(i_{\sigma(l+2)-1})})\cdots g(X_{\sigma(k)-1}^{(i_{\sigma(k)-1})})g(X_{\sigma(k+2)-1}^{(i_{\sigma(k+2)-1})})\cdots g(X_{\sigma(n+2)-1}^{(i_{\sigma(n+2)-1})}).
\end{aligned}
$$

For convenience we introduce the following notation: Since we assume that $1 \le i_1 \le \cdots \le i_{n+1} \le j$ and that there are at least two different numbers in any three of $i_1, \dots, i_{n+1}$, for any given $k$, $1 \le k \le n+1$, there are at most two of $i_1, \dots, i_{n+1}$ including $i_k$ equal to $i_k$ and if there are two of $i_1, \dots, i_{n+1}$ equal to $i_k$, these two must be successive. Let $j_1, \dots, j_q$ satisfying $1 \le j_1 < \cdots < j_q \le j$ be the distinct numbers in $\{i_1, \dots, i_{n+1}\}$. For any $m$, $1 \le m \le q$, if $j_m$ is equal to only one number $i_k$ in $\{i_1, \dots, i_{n+1}\}$, we let $g^{(j_m),1} = g(X_k^{(i_k)})$ and $g^{(i),2}$ the identity operator on $V^{\otimes j}$; if $j_m = i_k = i_{k+1}$ , we let $g^{(j_m),1} = g(X_k^{(i_k)})$ and $g^{(j_m),2} = g(X_{k+1}^{(i_{k+1})})$. we also define another bracket of two operators $A$ and $B$ of homogeneous fermion numbers by

$$[A, B]_1 = AB - (-1)^{(|A||B|-1)}BA. \tag{3.33}$$

Then using (3.13), the $k$-th term in the first sum of the right hand side of (3.32) is equal to: (i)

$$(-1)^{k+1}X_k^{(i_k)}\nu(P)\frac{[g^{(j_1),1}, g^{(j_1),2}]_1}{2}\cdots\Big(\frac{[g^{(j_m),1}, g^{(j_m),2}]_1}{2}\Big)^{\hat{}}\cdots\frac{[g^{(j_q),1}, g^{(j_q),2}]_1}{2} \tag{3.34}$$

(where $\hat{}$ means that the factor is omitted) if $i_k$ is the only number in $\{i_1, \dots, i_{n+1}\}$

equal to $j_m$; (ii)

$$(3.35)\qquad (-1)^{k+1}X_k^{(i_k)}\nu(P)\frac{[g^{(j_1),1},g^{(j_1),2}]_1}{2}\cdots g^{(j_m),2}\cdots\frac{[g^{(j_q),1},g^{(j_q),2}]_1}{2}$$

if $j_m = i_k = i_{k+1}$; (iii)

$$(3.36)\qquad (-1)^{k+1}X_k^{(i_k)}\nu(P)\frac{[g^{(j_1),1},g^{(j_1),2}]_1}{2}\cdots g^{(j_m),1}\cdots\frac{[g^{(j_q),1},g^{(j_q),2}]_1}{2}$$

if $j_m = i_{k-1} = i_k$. Since $X_l^{(i_l)}$, $l = 1,\dots,n+1$, are equal to either $-a_{i_l}^{-1}\frac{\partial}{\partial z_{i_l}}$ or $-a_{i_l}\frac{\partial}{\partial a_{i_l}}$, $g^{(j_r),1}$ and $g^{(j_r),2}$, $r = 1,\dots,q$, are constant as functions on $\hat{K}(j)$ by the definitions. Therefore, using (3.18), we have: (i)

$$(3.37)\quad X_k^{(i_k)}\nu(P)\frac{[g^{(j_1),1},g^{(j_1),2}]_1}{2}\cdots\left(\frac{[g^{(j_m),1},g^{(j_m),2}]_1}{2}\right)\hat{}\cdots\frac{[g^{(j_q),1},g^{(j_q),2}]_1}{2}$$
$$= \nu(P)L(X_k^{(i_k)})\frac{[g^{(j_1),1},g^{(j_m),2}]_1}{2}\cdots\left(\frac{[g^{(j_m),1},g^{(j_m),2}]_1}{2}\right)\hat{}\cdots\frac{[g^{(j_q),1},g^{(j_q),2}]_1}{2}$$

if $i_k$ is the only number in $\{i_1,\dots,i_{n+1}\}$ equal to $j_m$; (ii)

$$(3.38)\quad X_k^{(i_k)}\nu(P)\frac{[g^{(j_1),1},g^{(j_1),2}]_1}{2}\cdots g^{(j_m),2}\cdots\frac{[g^{(j_q),1},g^{(j_q),2}]_1}{2}$$
$$= \nu(P)L(X_k^{(i_k)})\frac{[g^{(j_1),1},g^{(j_1),2}]_1}{2}\cdots g^{(j_m),2}\cdots\frac{[g^{(j_q),1},g^{(j_q),2}]_1}{2}$$

if $j_m = i_k = i_{k+1}$; (iii)

$$(3.39)\quad X_k^{(i_k)}\nu(P)\frac{[g^{(j_1),1},g^{(j_1),2}]_1}{2}\cdots g^{(j_m),1}\cdots\frac{[g^{(j_q),1},g^{(j_q),2}]_1}{2}$$
$$= \nu(P)L(X_k^{(i_k)})\frac{[g^{(j_1),1},g^{(j_1),2}]_1}{2}\cdots g^{(j_m),1}\cdots\frac{[g^{(j_q),1},g^{(j_q),2}]_1}{2}$$

if $j_m = i_{k-1} = i_k$. Using (3.15) we see that the $k$-th term of the right hand side of (3.32) is equal to

$$(3.40)\qquad (-1)^{k+1}\nu(P)\frac{[g^{(j_1),1},g^{(j_1),2}]_1}{2}\cdots L(X_k^{(i_k)})\cdots\frac{[g^{(j_q),1},g^{(j_q),2}]_1}{2}$$

in the case (i), is equal to

$$(3.41)\qquad (-1)^{k+1}\nu(P)\frac{[g^{(j_1),1},g^{(j_1),2}]_1}{2}\cdots L(X_k^{(i_k)})g^{(j_m),2}\cdots\frac{[g^{(j_q),1},g^{(j_q),2}]_1}{2}$$

In the case (ii) and is equal to

$$(3.42)\qquad (-1)^{k+1}\nu(P)\frac{[g^{(j_1),1},g^{(j_1),2}]_1}{2}\cdots L(X_k^{(i_k)})g^{(j_m),1}\cdots\frac{[g^{(j_q),1},g^{(j_q),2}]_1}{2}$$

in the case (iii). Using the axioms (2.15) and (2.16), we have

$$L(X_k^{(i_k)}) = [Q_{V^{\otimes j}}, g(X_k^{(i_k)})] = Q_{V^{\otimes j}} g(X_k^{(i_k)}) + g(X_k^{(i_k)}) Q_{V^{\otimes j}} \tag{3.43}$$

where $Q_{V^{\otimes j}} : V^{\otimes j} \to V^{\otimes j}$ is defined by (3.24). Thus the $k$-th term of the right hand side of (3.32) is equal to

(3.44)

$$\begin{aligned}
&(-1)^{k+1}\nu(P)\frac{[g^{(j_1),1}, g^{(j_1),2}]_1}{2}\cdots Q_{V^{\otimes j}} g(X_k^{(i_k)})\cdots \frac{[g^{(j_q),1}, g^{(j_q),2}]_1}{2}\\
&\quad\quad + (-1)^{k+1}\nu(P)\frac{[g^{(j_1),1}, g^{(j_1),2}]_1}{2}\cdots g(X_k^{(i_k)}) Q_{V^{\otimes j}}\cdots \frac{[g^{(j_q),1}, g^{(j_q),2}]_1}{2}\\
&= (-1)^{k+1}\nu(P)\frac{[g^{(j_1),1}, g^{(j_1),2}]_1}{2}\cdots Q_{V^{\otimes j}} g^{(j_m),1}\cdots \frac{[g^{(j_q),1}, g^{(j_q),2}]_1}{2}\\
&\quad + (-1)^{k+1}\nu(P)\frac{[g^{(j_1),1}, g^{(j_1),2}]_1}{2}\cdots g^{(j_m),1} Q_{V^{\otimes j}}\cdots \frac{[g^{(j_q),1}, g^{(j_q),2}]_1}{2}\\
&= (-1)^{k+1}\nu(P)\frac{[g^{(j_1),1}, g^{(j_1),2}]_1}{2}\cdots Q_{V^{\otimes j}} \frac{[g^{(j_m),1}, g^{(j_m),2}]_1}{2}\cdots \frac{[g^{(j_q),1}, g^{(j_q),2}]_1}{2}\\
&\quad + (-1)^{k+1}\nu(P)\frac{[g^{(j_1),1}, g^{(j_1),2}]_1}{2}\cdots \frac{[g^{(j_m),1}, g^{(j_m),2}]_1}{2} Q_{V^{\otimes j}}\cdots \frac{[g^{(j_q),1}, g^{(j_q),2}]_1}{2}.
\end{aligned}$$

in the case (i), is equal to

(3.45)

$$\begin{aligned}
&(-1)^{k+1}\nu(P)\frac{[g^{(j_1),1}, g^{(j_1),2}]_1}{2}\cdots Q_{V^{\otimes j}} g(X_k^{(i_k)}) g^{(j_m),2}\cdots \frac{[g^{(j_q),1}, g^{(j_q),2}]_1}{2}\\
&\quad\quad + (-1)^{k+1}\nu(P)\frac{[g^{(j_1),1}, g^{(j_1),2}]_1}{2}\cdots g(X_k^{(i_k)}) Q_{V^{\otimes j}} g^{(j_m),2}\cdots \frac{[g^{(j_q),1}, g^{(j_q),2}]_1}{2}\\
&= (-1)^{k+1}\nu(P)\frac{[g^{(j_1),1}, g^{(j_1),2}]_1}{2}\cdots Q_{V^{\otimes j}} g^{(j_m),1} g^{(j_m),2}\cdots \frac{[g^{(j_q),1}, g^{(j_q),2}]_1}{2}\\
&\quad + (-1)^{k+1}\nu(P)\frac{[g^{(j_1),1}, g^{(j_1),2}]_1}{2}\cdots g^{(j_m),1} Q_{V^{\otimes j}} g^{(j_m),2}\cdots \frac{[g^{(j_q),1}, g^{(j_q),2}]_1}{2}
\end{aligned}$$

in the case (ii) and is equal to

(3.46)
$$\begin{aligned}&(-1)^{k+1}\nu(P)\frac{[g^{(j_1),1},g^{(j_1),2}]_1}{2}\cdots[L(X_k^{(i_k)}),g^{(j_m),1}]\cdots\frac{[g^{(j_q),1},g^{(j_q),2}]_1}{2}\\&\qquad+(-1)^{k+1}\nu(P)\frac{[g^{(j_1),1},g^{(j_1),2}]_1}{2}\cdots g^{(j_m),1}L(X_k^{(i_k)})\cdots\frac{[g^{(j_q),1},g^{(j_q),2}]_1}{2}\\=&(-1)^{k+1}\nu(P)\frac{[g^{(j_1),1},g^{(j_1),2}]_1}{2}\cdots[L(X_k^{(i_k)}),g^{(j_m),1}]\cdots\frac{[g^{(j_q),1},g^{(j_q),2}]_1}{2}\\&+(-1)^{k+1}\nu(P)\frac{[g^{(j_1),1},g^{(j_1),2}]_1}{2}\cdots g^{(j_m),1}Q_{V^{\otimes j}}g^{(j_m),2}\cdots\frac{[g^{(j_q),1},g^{(j_q),2}]_1}{2}\\&+(-1)^{k+1}\nu(P)\frac{[g^{(j_1),1},g^{(j_1),2}]_1}{2}\cdots g^{(j_m),1}g^{(j_m),2}Q_{V^{\otimes j}}\cdots\frac{[g^{(j_q),1},g^{(j_q),2}]_1}{2}\end{aligned}$$

in the case (iii).

We prove that

(3.47)
$$Q_{V^{\otimes j}}g^{(j_m),1}g^{(j_m),2}-g^{(j_m),1}g^{(j_m),2}Q_{V^{\otimes j}}=Q_{V^{\otimes j}}\frac{[g^{(j_m),1},g^{(j_m),2}]_1}{2}-\frac{[g^{(j_m),1},g^{(j_m),2}]_1}{2}Q_{V^{\otimes j}}$$

for any $m$ such that $i_{k-1}=i_k=j_m$. Using (3.43), we have

(3.48)
$$\begin{aligned}&Q_{V^{\otimes j}}g^{(j_m),1}g^{(j_m),2}-g^{(j_m),1}g^{(j_m),2}Q_{V^{\otimes j}}\\&\quad=Q_{V^{\otimes j}}[g^{(j_m),1},g^{(j_m),2}]_1+Q_{V^{\otimes j}}g^{(j_m),2}g^{(j_m),1}\\&\qquad-g^{(j_m),2}g^{(j_m),1}Q_{V^{\otimes j}}-[g^{(j_m),1},g^{(j_m),2}]_1Q_{V^{\otimes j}}\\&\quad=Q_{V^{\otimes j}}[g^{(j_m),1},g^{(j_m),2}]_1-[g^{(j_m),1},g^{(j_m),2}]_1Q_{V^{\otimes j}}\\&\qquad+Q_{V^{\otimes j}}g(X_k^{(i_k)})g(X_{k-1}^{(i_{k-1})})-g(X_k^{(i_k)})g(X_{k-1}^{(i_{k-1})})Q_{V^{\otimes j}}\\&\quad=Q_{V^{\otimes j}}[g^{(j_m),1},g^{(j_m),2}]_1-[g^{(j_m),1},g^{(j_m),2}]_1Q_{V^{\otimes j}}\\&\qquad-g(X_k^{(i_k)})Q_{V^{\otimes j}}g(X_{k-1}^{(i_{k-1})})+L(X_k^{(i_k)})g(X_{k-1}^{(i_{k-1})})-g(X_k^{(i_k)})g(X_{k-1}^{(i_{k-1})})Q_{V^{\otimes j}}\\&\quad=Q_{V^{\otimes j}}[g^{(j_m),1},g^{(j_m),2}]_1-[g^{(j_m),1},g^{(j_m),2}]_1Q_{V^{\otimes j}}+g(X_k^{(i_k)})g(X_{k-1}^{(i_{k-1})})Q_{V^{\otimes j}}\\&\qquad-g(X_k^{(i_k)})L(X_{k-1}^{(i_{k-1})})+L(X_k^{(i_k)})g(X_{k-1}^{(i_{k-1})})-g(X_k^{(i_k)})g(X_{k-1}^{(i_{k-1})})Q_{V^{\otimes j}}\\&\quad=Q_{V^{\otimes j}}[g^{(j_m),1},g^{(j_m),2}]_1-[g^{(j_m),1},g^{(j_m),2}]_1Q_{V^{\otimes j}}\\&\qquad-g(X_k^{(i_k)})L(X_{k-1}^{(i_{k-1})})+L(X_k^{(i_k)})g(X_{k-1}^{(i_{k-1})}).\end{aligned}$$

Since

(3.49) $g([X_{k-1}^{(i_{k-1})},X_k^{(i_k)}])$

$$= -[L(X_k^{(i_k)}), g(X_{k-1}^{(i_{k-1})})] = -L(X_k^{(i_k)})g(X_{k-1}^{(i_{k-1})}) + g(X_{k-1}^{(i_{k-1})})L(X_k^{(i_k)}),$$

$$\begin{aligned}
(3.50)\quad & g([X_{k-1}^{(i_{k-1})}, X_k^{(i_k)}]) \\
&= [L(X_{k-1}^{(i_{k-1})}), g(X_k^{(i_k)})] = L(X_{k-1}^{(i_{k-1})})g(X_k^{(i_k)}) - g(X_k^{(i_k)})L(X_{k-1}^{(i_{k-1})}),
\end{aligned}$$

we have

(3.51)
$$-L(X_k^{(i_k)})g(X_{k-1}^{(i_{k-1})}) + g(X_{k-1}^{(i_{k-1})})L(X_k^{(i_k)}) = L(X_{k-1}^{(i_{k-1})})g(X_k^{(i_k)}) - g(X_k^{(i_k)})L(X_{k-1}^{(i_{k-1})})$$

or

(3.52)
$$-g(X_k^{(i_k)})L(X_{k-1}^{(i_{k-1})}) + L(X_k^{(i_k)})g(X_{k-1}^{(i_{k-1})}) = g(X_{k-1}^{(i_{k-1})})L(X_k^{(i_k)}) - L(X_{k-1}^{(i_{k-1})})g(X_k^{(i_k)}).$$

Thus

$$\begin{aligned}
& - g(X_k^{(i_k)})L(X_{k-1}^{(i_{k-1})}) + L(X_k^{(i_k)})g(X_{k-1}^{(i_{k-1})}) \\
(3.53)\quad &= \frac{1}{2}(-g(X_k^{(i_k)})L(X_{k-1}^{(i_{k-1})}) + L(X_k^{(i_k)})g(X_{k-1}^{(i_{k-1})}) \\
&\qquad + g(X_{k-1}^{(i_{k-1})})L(X_k^{(i_k)}) - L(X_{k-1}^{(i_{k-1})})g(X_k^{(i_k)}))
\end{aligned}$$

Using (3.43), the right hand side of (3.53) becomes

$$\begin{aligned}
& \frac{1}{2}( - g(X_k^{(i_k)})(Q_{V^{\otimes j}}g(X_{k-1}^{(i_{k-1})}) + g(X_{k-1}^{(i_{k-1})})Q_{V^{\otimes j}}) \\
& \quad + (Q_{V^{\otimes j}}g(X_k^{(i_k)}) + g(X_k^{(i_k)})Q_{V^{\otimes j}})g(X_{k-1}^{(i_{k-1})}) \\
& \quad + g(X_{k-1}^{(i_{k-1})})(Q_{V^{\otimes j}}g(X_k^{(i_k)}) + g(X_k^{(i_k)})Q_{V^{\otimes j}}) \\
& \quad - (Q_{V^{\otimes j}}g(X_{k-1}^{(i_{k-1})}) + g(X_{k-1}^{(i_{k-1})})Q_{V^{\otimes j}})g(X_k^{(i_k)})) \\
(3.54)\quad & = \frac{1}{2}(-g(X_k^{(i_k)})g(X_{k-1}^{(i_{k-1})})Q_{V^{\otimes j}} + Q_{V^{\otimes j}}g(X_k^{(i_k)})g(X_{k-1}^{(i_{k-1})}) \\
& \qquad + g(X_{k-1}^{(i_{k-1})})g(X_k^{(i_k)})Q_{V^{\otimes j}} - Q_{V^{\otimes j}}g(X_{k-1}^{(i_{k-1})})g(X_k^{(i_k)})) \\
& = -Q_{V^{\otimes j}}\frac{[g(X_{k-1}^{(i_{k-1})}), g(X_k^{(i_k)})]_1}{2} + \frac{[g(X_{k-1}^{(i_{k-1})}), g(X_k^{(i_k)})]_1}{2}Q_{V^{\otimes j}} \\
& = -Q_{V^{\otimes j}}\frac{[g^{(j_m),1}, g^{(j_m),2}]_1}{2} + \frac{[g^{(j_m),1}, g^{(j_m),2}]_1}{2}Q_{V^{\otimes j}}.
\end{aligned}$$

Substituting (3.54) into the right hand side of (3.48), we obtain (3.47).

Now we calculate the second sum in the right hand side of (3.32). Since $X_l^{(i_l)}$, $l = 1, \dots, n+1$, are equal to either $-a_{i_l}^{-1}\frac{\partial}{\partial z_{i_l}}$ or $-a_{i_l}\frac{\partial}{\partial a_{i_l}}$,

$$g([X_l^{(i_l)}, X_k^{(i_k)}]) = 0 \tag{3.55}$$

if $i_l \neq i_k$. Therefore we need to consider only those terms in the sum such that $i_l = i_k$, that is, $l = k-1$ and $i_{k-1} = i_k = j_m$ for some $m$, $1 \le m \le q$. From (3.15), we have

$$\begin{aligned} g([X_{k-1}^{(i_{k-1})}, X_k^{(i_k)}]) &= -g([X_k^{(i_k)}, X_{k-1}^{(i_{k-1})}]) \\ &= -[L(X_k^{(i_k)}), g(X_{k-1}^{(i_{k-1})})] \\ &= -[L(X_k^{(i_k)}), g^{(j_m),1}]. \end{aligned} \tag{3.56}$$

Thus these terms are of the form

(3.57)
$$\begin{aligned} &-\nu(P)g([X_{k-1}^{(i_{k-1})}, X_k^{(i_k)}])\frac{[g^{(j_1),1}, g^{(j_1),2}]_1}{2}\cdots\Big(\frac{[g^{(j_m),1}, g^{(j_m),2}]_1}{2}\Big)^{\wedge}\cdots\frac{[g^{(j_{n+1}),1}, g^{(j_{n+1}),2}]_1}{2} \\ &\quad = (-1)^{k-1}\nu(P)\frac{[g^{(j_1),1}, g^{(j_1),2}]_1}{2}\cdots\frac{[g^{(j_{m-1}),1}, g^{(j_{m-1}),2}]_1}{2}g([X_k^{(i_{k-1})}, X_l^{(i_k)}])\cdot \\ &\qquad\qquad \cdot\frac{[g^{(j_{m+1}),1}, g^{(j_{m+1}),2}]_1}{2}\cdots\frac{[g^{(j_{n+1}),1}, g^{(j_{n+1}),2}]_1}{2} \\ &\quad = (-1)^{k}\nu(P)\frac{[g^{(j_1),1}, g^{(j_1),2}]_1}{2}\cdots\frac{[g^{(j_{m-1}),1}, g^{(j_{m-1}),2}]_1}{2}[L(X_k^{(i_k)}), g^{(j_m),1}]\cdot \\ &\qquad\qquad \cdot\frac{[g^{(j_{m+1}),1}, g^{(j_{m+1}),2}]_1}{2}\cdots\frac{[g^{(j_{n+1}),1}, g^{(j_{n+1}),2}]_1}{2} \end{aligned}$$

Substituting (3.44) — (3.46) and (3.57) into (3.32) and using (3.47), we obtain

$$
\begin{aligned}
&(d\omega_{j,n})(X_1^{(i_1)}, \dots, X_{n+1}^{(i_{n+1})})\\
&= \nu(P)Q_{V^{\otimes j}}\frac{[g^{(j_1),1}, g^{(j_1),2}]_1}{2}\cdots\frac{[g^{(j_{n+1}),1}, g^{(j_{n+1}),2}]_1}{2}\\
&\quad + (-1)^{n+2}\nu(P)\frac{[g^{(j_1),1}, g^{(j_1),2}]_1}{2}\cdots\frac{[g^{(j_{n+1}),1}, g^{(j_{n+1}),2}]_1}{2}Q_{V^{\otimes j}}\\
&= \nu(P)Q_{V^{\otimes j}}\frac{1}{(n+1)!}\sum_{\sigma\in S_{n+1}}(-1)^{\mathrm{sgn}\sigma}g(X_{\sigma(1)}^{(i_{\sigma(1)})})\cdots g(X_{\sigma(n+1)}^{(i_{\sigma(n+1)})})\\
&\quad + (-1)^{n+2}\nu(P)\frac{1}{(n+1)!}\sum_{\sigma\in S_{n+1}}(-1)^{\mathrm{sgn}\sigma}g(X_{\sigma(1)}^{(i_{\sigma(1)})})\cdots g(X_{\sigma(n+1)}^{(i_{\sigma(n+1)})})Q_{V^{\otimes j}}\\
&= Q\nu(P)\frac{1}{(n+1)!}\sum_{\sigma\in S_{n+1}}(-1)^{\mathrm{sgn}\sigma}g(X_{\sigma(1)}^{(i_{\sigma(1)})})\cdots g(X_{\sigma(n+1)}^{(i_{\sigma(n+1)})})\\
&\quad + (-1)^{n+2}\nu(P)\frac{1}{(n+1)!}\sum_{\sigma\in S_{n+1}}(-1)^{\mathrm{sgn}\sigma}g(X_{\sigma(1)}^{(i_{\sigma(1)})})\cdots g(X_{\sigma(n+1)}^{(i_{\sigma(n+1)})})Q_{V^{\otimes j}}\\
&= Q_{\mathcal{H}_{V,0}}(\omega_{j,n+1}(X_1^{(i_1)}, \dots, X_{n+1}^{(i_{n+1})})). \quad \square
\end{aligned} \tag{3.58}
$$

The third property describes how the partial operad $\hat{K}$ (or $\overline{K}$) acts on $\omega_j$, $j \in \mathbb{N}$. In fact it is this property which characterizes the three classes of topological vertex algebras. Let $P_0 \in \hat{K}(k)$, $P_1 \in \hat{K}(j_1), \dots, P_k \in \hat{K}(j_k)$. Assume that $\gamma_{\hat{K}}(P_0; P_1, \dots, P_k)$ exists. Since there is a natual isomorphism from $T_{P_0}(\hat{K}(k)) \oplus T_{P_1}(\hat{K}(j_1)) \oplus \cdots \oplus T_{P_k}(\hat{K}(j_k))$ to $T_{(P_0,\dots,P_k)}(\hat{K}(k) \times \hat{K}(j_1) \times \cdots \times \hat{K}(j_k))$, the tangent spaces $T_{P_0}(\hat{K}(k))$, $T_{P_1}(\hat{K}(j_1)), \dots, T_{P_k}(\hat{K}(j_k))$ can all be embedded in $T_{(P_0,\dots,P_k)}(\hat{K}(k) \times \hat{K}(j_1) \times \cdots \times \hat{K}(j_k))$. We denote these embeddings by $e^{(0)}, \dots, e^{(k)}$, respectively. For convenience we sometimes use the notation $j_0 = k$. Let $X_{li} \in T_{P_l}(\hat{K}(j_l))$, $i = 1, \dots, n_l$, $l = 0, \dots, k$. Then $(\gamma_{\hat{K}})_*(e_*^{(l)}(X_{li}))$, $i = 1, \dots, n_l$, $l = 0, \dots, k$, are elements of $T_{\gamma_{\hat{K}}(P_0;P_1,\dots,P_k)}(\hat{K}(j_1 + \cdots + j_k))$. We also denote the space of holomorphic functions on a manifold $M$ by $\mathrm{Hol}(M)$.

The manifold $\hat{K}(k) \times \hat{K}(j_1) \times \cdots \times \hat{K}(j_k)$ has natural foliations whose leaves are obtained by fixing certain punctures and the corresponding local coordinates of the projection images of its elements to $\hat{K}(k)$, $\hat{K}(j_1), \cdots, \hat{K}(j_k)$. We call these foliations *coordinate foliations*. We also have foliations of $\hat{K}(k) \times \hat{K}(j_1) \times \cdots \times \hat{K}(j_k)$ whose leaves are obtained by fixing certain punctures and the corresponding local coordinates of the images of its elements under the map $\gamma_{\hat{K}} : \hat{K}(k) \times \hat{K}(j_1) \times \cdots \times \hat{K}(j_k) \to \hat{K}(j_1 + \cdots + j_k)$. We call these foliations *pull-back coordinate foliations.*

**Proposition 3.3.** *Let $V$ be a strong topological vertex algebra and $\gamma_{\mathcal{H}_{V,0}}$ be the substitution map (composition map) for the endomorphism partial pseudo-operad $\mathcal{H}_{V,0}$.*

*We have*

$$\omega_{j_1+\cdots+j_k,n_0+\cdots+n_k}((\gamma_{\hat{K}})_*(e_*^{(0)}(X_{01})),\dots,(\gamma_{\hat{K}})_*(e_*^{(k)}(X_{kn_k})))\big|_{\gamma_{\hat{K}}(P_0;P_1,\dots,P_k)}$$
$$= \gamma_{\mathcal{H}_{V,0}}(\omega_{k,n_0}(e_*^{(0)}(X_{01}),\dots,e_*^{(0)}(X_{0n_0}))\big|_{P_0};\omega_{j_1,n_1}(e_*^{(1)}(X_{11}),\dots,e_*^{(0)}(X_{1j_1}))\big|_{P_1},$$
$$\dots,\omega_{j_k,n_k}(e_*^{(k)}(X_{k1}),\dots,e_*^{(k)}(X_{kn_k}))\big|_{P_k}) \tag{3.59}$$

*for $P_l\in\hat{K}$, $l=0,\dots,k$ and $X_{li}\in T_{P_l}(\hat{K}(j_l))$, $i=1,\dots,n_l$, $l=0,\dots,k$. If $V$ is a topological vertex algebra, then there exist maps*

$$h,\alpha:(\Gamma(T(\hat{K}(k)\times\hat{K}(j_1)\times\cdots\times\hat{K}(j_k))))^{\otimes(n_0+\cdots+n_k)}$$
$$\to\mathcal{H}_{V,0}(j_1+\cdots+j_k)\otimes\mathrm{Hol}(\hat{K}(k)\times\hat{K}(j_1)\times\cdots\times\hat{K}(j_k))$$

*where $\alpha$ is a sum of maps which are holomorphic forms when restricted to certain tensor factors of*

$$(\Gamma(T(\hat{K}(k)\times\hat{K}(j_1)\times\cdots\times\hat{K}(j_k))))^{\otimes(n_0+\cdots+n_k)}$$

*and are exact forms when these holomorphic forms are restricted to leaves of certain coordinate foliations or pull-back coordinate foliations of $\hat{K}(k)\times\hat{K}(j_1)\times\cdots\times\hat{K}(j_k)$, such that*

$$\omega_{j_1+\cdots+j_k,n_0+\cdots+n_k}((\gamma_{\hat{K}})_*(e_*^{(0)}(X_{01})),\dots,(\gamma_{\hat{K}})_*(e_*^{(k)}(X_{kn_k})))\big|_{\gamma_{\hat{K}}(P_0;P_1,\dots,P_k)}$$
$$= \gamma_{\mathcal{H}_{V,0}}(\omega_{k,n_0}(e_*^{(0)}(X_{01}),\dots,e_*^{(0)}(X_{0n_0}))\big|_{P_0};\omega_{j_1,n_1}(e_*^{(1)}(X_{11}),\dots,e_*^{(0)}(X_{1j_1}))\big|_{P_1},$$
$$\dots,\omega_{j_k,n_k}(e_*^{(k)}(X_{k1}),\dots,e_*^{(k)}(X_{kn_k}))\big|_{P_k})$$
$$+Q_{\mathcal{H}_{V,0}}(h((\gamma_{\hat{K}})_*(e_*^{(0)}(X_{01})),\dots,(\gamma_{\hat{K}})_*(e_*^{(k)}(X_{kn_k})))\big|_{(P_0,\dots,P_k)})$$
$$+\alpha((\gamma_{\hat{K}})_*(e_*^{(0)}(X_{01})),\dots,(\gamma_{\hat{K}})_*(e_*^{(k)}(X_{kn_k})))\big|_{(P_0,\dots,P_k)}. \tag{3.60}$$

*Moreover,*

$$h=0\quad\text{when } n_0=0 \text{ or } n_1=\cdots=n_k=0 \tag{3.61}$$

*and*

$$\alpha=0\quad\text{when } n_0,n_1,\dots,n_k=0,1. \tag{3.62}$$

*If $V$ is a weak topological vertex algebra, (3.59) holds when $n_0=0$ or $n_1=\cdots=n_k=0$ and (3.60) and (3.62) hold for $P_l\in\overline{K}$, $l=0,\dots,k$ and $X_{li}\in T_{P_l}(\overline{K}(j_l))$, $i=1,\dots,n_l$, $l=0,\dots,k$.*

*Proof.* We first prove (3.59) for strong topological vertex algebras. For convenience we use the notations $Y_{li}=e_*^{(l)}(X_{li})$, $i=1,\dots,n_l$, $l=0,\dots,k$. Also in the proof below, we regard $P_0, \dots, P_k$ as variables and omit the the symbols $\big|_{P_0}, \dots, \big|_{P_k}, \big|_{\gamma_{\hat{K}}(P_0; P_1, \dots, P_k)}$ and $\big|_{(P_0, \dots, P_k)}$. By the definition of $\omega_{j_1+\cdots+j_k, n_0+\cdots+n_k}$, we have

$$\begin{aligned}\omega_{j_1+\cdots+j_k, n_0+\cdots+n_k}&((\gamma_{\hat{K}})_*(Y_{01}), \dots, (\gamma_{\hat{K}})_*(Y_{kn_k}))\\ &= \nu(\gamma_{\hat{K}}(P_0; P_1, \dots, P_k))g((\gamma_{\hat{K}})_*(Y_{01})) \wedge \cdots \wedge g((\gamma_{\hat{K}})_*(Y_{kn_k})).\end{aligned} \tag{3.63}$$

Since for strong topological vertex algebras, $g((\gamma_{\hat{K}})_*((Y_{ls}))$ and $g((\gamma_{\hat{K}})_*((Y_{mt}))$ anti-commute with each other for $s = 1, \dots, n_l$, $t = 1, \dots, n_m$, $l, m = 0, \dots, k$, the right hand side of ( ) can be written as

$$\begin{aligned}\nu(\gamma_{\hat{K}}(P_0; P_1, \dots, P_k))&(g((\gamma_{\hat{K}})_*(Y_{01})) \wedge \cdots \wedge g((\gamma_{\hat{K}})_*(Y_{0n_0})))\\ &\cdots (g((\gamma_{\hat{K}})_*(Y_{1n_k})) \wedge \cdots \wedge g((\gamma_{\hat{K}})_*(Y_{kn_k})))\end{aligned} \tag{3.64}$$

On the other hand,

$$\begin{aligned}&\gamma_{\mathcal{H}_{V,W}}(\omega_{k,n_0}(Y_{01}, \dots, Y_{0n_0}); \omega_{j_1,n_1}(Y_{11}, \dots, Y_{1n_1}), \dots, \omega_{j_k,n_k}(Y_{k1}, \dots, Y_{kn_k}))\\ &\quad = \gamma_{\mathcal{H}_{V,W}}(\nu(P_0)g(Y_{01}) \wedge \cdots \wedge g(Y_{0n_0});\\ &\qquad\qquad \nu(P_1)g(Y_{11}) \wedge \cdots \wedge g(Y_{1n_1}), \dots, \nu(P_k)g(Y_{k1}) \wedge \cdots \wedge g(Y_{kn_k})).\end{aligned} \tag{3.65}$$

To prove that (3.64) is equal the right hand side of (3.65), we need only to prove

$$\begin{aligned}&\nu(\gamma_{\hat{K}}(P_0; P_1, \dots, P_k))(g((\gamma_{\hat{K}})_*(Y_{01})) \cdots g((\gamma_{\hat{K}})_*(Y_{0n_0})))\\ &\qquad\qquad \cdots (g((\gamma_{\hat{K}})_*(Y_{1n_k})) \cdots g((\gamma_{\hat{K}})_*(Y_{kn_k})))\\ &\quad = \gamma_{\mathcal{H}_{V,W}}(\nu(P_0)g(Y_{01}) \cdots g(Y_{0n_0}); \nu(P_1)g(Y_{11}) \cdots g(Y_{1n_1}),\\ &\qquad\qquad \dots, \nu(P_k)g(Y_{k1}) \cdots g(Y_{kn_k}))\end{aligned} \tag{3.66}$$

To prove (3.66), we first note that the bracket formulas of $g(0)$, $g(-1)$ with vertex operators and with $L(0)$, $L(-1)$ are the same as those of $L(0)$, $L(-1)$ with vertex operators and with $L(0)$, $L(-1)$ themselves, except suitable signs coming from the difference between the fermion numbers of $g(0)$, $g(-1)$ and of $L(0)$, $L(-1)$. Therefore if we can prove

$$\begin{aligned}\nu(\gamma_{\hat{K}}(P_0; P_1, \dots, P_k))&L((\gamma_{\hat{K}})_*(Y_{01})) \cdots L((\gamma_{\hat{K}})_*(Y_{kn_k}))\\ &= \gamma_{\mathcal{H}_{V,W}}(\nu(P_0)L(Y_{01}) \cdots L(Y_{0n_0});\\ &\qquad \nu(P_1)L(Y_{11}) \cdots L(Y_{1n_1}), \dots, \nu(P_k)L(Y_{k1}) \cdots L(Y_{kn_k}))\end{aligned} \tag{3.67}$$

without using the bracket formulas for $L(0)$ and $L(-1)$, (3.66) will be a consequence. To prove (3.67), we first use (3.18) and the fact that $\nu$ is a morphism of partial pseudo-operads to obtain

$$\nu(\gamma_{\hat{K}}(P_0; P_1, \dots, P_k))L((\gamma_{\hat{K}})_*(Y_{01})) \cdots L((\gamma_{\hat{K}})_*(Y_{kn_k})) \tag{3.68}$$

$$= (\gamma_{\hat{K}})_*(Y_{01}) \cdots (\gamma_{\hat{K}})_*(Y_{kn_k}) \nu(\gamma_{\hat{K}}(P_0; P_1, \dots, P_k))$$
$$= (\gamma_{\hat{K}})_*(Y_{01}) \cdots (\gamma_{\hat{K}})_*(Y_{kn_k}) \gamma_{\mathcal{H}_{V,W}}(\nu(P_0); \nu(P_1), \dots, \nu(P_k)).$$

Note that as differential operators $(\gamma_{\hat{K}})_*(Y_{li})$'s are the same as $Y_{li}$'s. We denote these differential operators as $\partial_{li}$. Then

$$\begin{aligned}
&(\gamma_{\hat{K}})_*(Y_{01}) \cdots (\gamma_{\hat{K}})_*(Y_{kn_k}) \gamma_{\mathcal{H}_{V,W}}(\nu(P_0); \nu(P_1), \dots, \nu(P_k))\\
&\quad = \partial_{01} \cdots \partial_{kn_k} \gamma_{\mathcal{H}_{V,W}}(\nu(P_0); \nu(P_1), \dots, \nu(P_k))\\
&\quad = \gamma_{\mathcal{H}_{V,W}}(\partial_{01} \cdots \partial_{0n_0} \nu(P_0); \partial_{11} \cdots \partial_{1n_1} \nu(P_1), \dots, \partial_{k1} \cdots \partial_{kn_k} \nu(P_k))\\
&\quad = \gamma_{\mathcal{H}_{V,W}}(\nu(P_0) L(Y_{01}) \cdots L(Y_{0n_0});\\
&\qquad\qquad \nu(P_1) L(Y_{11}) \cdots L(Y_{1n_1}), \dots, \nu(P_k) L(Y_{k1}) \cdots L(Y_{kn_k}))
\end{aligned} \tag{3.69}$$

where we have used (3.18) and

$$\partial_{li} \nu(P_m) = 0, \quad i = 1, \dots, n_l, \quad l \neq m. \tag{3.70}$$

This proves (3.59). Note that the proof above also proves that (3.64) is equal to the right hand side of (3.65) even for weak topological vertex algebras.

To prove the case for topological vertex algebras, we first notice that from the definition that

$$(\gamma_{\hat{K}})_*(Y_{li}) \in \bigoplus_{j=j_1+\cdots+j_{l-1}+1}^{j_1+\cdots+j_l} \Gamma(T^{(j)}(\hat{K}(j_1 + \cdots + j_k))),$$

$i = 1, \dots, n_l$. Thus by (3.13) $g((\gamma_{\hat{K}})_*(Y_{li}))$ and $g((\gamma_{\hat{K}})_*(Y_{mi}))$ anti-commute with each other if $l \neq m$ and $l, m \neq 0$. We take $X_{li}$ to be in $\Gamma(T^{(p)}(\hat{K}(j_l)))$ for some $p$, $1 \leq p \leq n_l$. Then if there exists $l$ and $p$ such that more than two of $X_{li}$, $i = 1, \dots, n_l$, are in $\Gamma(T^{(p)}(\hat{K}(j_l)))$, both the left hand side and the first term in the right hand side of (3.60) are zero. Thus we can define the values of $h$ and $\alpha$ on these tangent vector fields to be zero. The formula (3.60) holds in these cases. Thus we can assume that for any $l$ and any $p$ there are at most two of $X_{li}$, $i = 1, \dots, n_l$, in $\Gamma(T^{(p)}(\hat{K}(j_l)))$. In these cases the left hand side of (3.60) can be written as a sum with suitable signs of terms of the form

$$\nu(\gamma_{\hat{K}}(P_0; P_1, \dots, P_k)) A^{(1)} \cdots A^{(j_1+\cdots+j_k)} \tag{3.71}$$

where $A^{(q)}$, $q = 1, \dots, j_1 + \cdots + j_k$, are operators on $V^{\otimes(j_1+\cdots+j_k)}$ equal to the skew-symmetrization of the product of at most four of $g(((\gamma_{\hat{K}})_*(Y_{li}))^{(q)})$ such that at most two of $X_{li}$ are in the set $\{X_{0i} | i = 1, \dots, n_0\}$ and at most two of $X_{li}$ are in the set $\{X_{li} | i = 1, \dots, n_l, l = 1, \dots, k\}$. We need the following lemma

**Lemma 3.4.** *For any $q$ satisfying $1 \leq q \leq j_1 + \cdots + j_k$, we can write $q = j_1 + \cdots j_{l-1} + i$ for some $l, i$ satisfying $1 \leq l \leq k$, $1 \leq i \leq j_l$, respectively. Then $A^{(q)}$ can be written as a sum of three terms: the first term is the product of the skew-symmetrization* *of the product of those $g(((\gamma_{\hat{K}})_*(Y_{ms}))^{(q)})$'s with $X_{ms}$'s in $\{X_{0i}|s = 1,\dots,n_0\}$ and the skew-symmetrization of the product of those $g(((\gamma_{\hat{K}})_*(Y_{ms}))^{(q)})$'s with $X_{ms}$'s in $\{X_{ls}|s = 1,\dots,n_l, l = 1,\dots,k\}$; the second term is $Q$-exact; and the third term satisfies the property that the composition of $\nu(\gamma_{\hat{K}}(P_0;P_1,\dots,P_k))$ with it is an exact form evaluated at those tangent vector fields defining $A^{(q)}$ on any leave of the foliation of $\hat{K}(k)\times\hat{K}(j_1)\times\cdots\times\hat{K}(j_k)$ obtained by fixing all punctures and local coordinates of elements of $\hat{K}(j_m)$, $m = 0,\dots,k$ except the the $l$-th ones of elements of $\hat{K}(k)$ and the $i$-th ones of elements of $\hat{K}(j_l)$.*

*Proof.* For simplicity, we prove only the case that $A^{(q)}$ is the skew-symmetrization of the product of three of $g(((\gamma_{\hat{K}})_*(Y_{ms}))^{(q)})$ with only one of them obtained from a tangent vector field on $\hat{K}(k)$. Let $P_m = (z_{m1},\dots,z_{mj_m}; a_{m1},\dots,a_{mj_m})$, $m = 0,\dots,k$. We assume that the corresponding three $X_{ms}$'s are $-a_{0l}\frac{\partial}{\partial a_{0l}}$, $-a_{li}^{-1}\frac{\partial}{\partial z_{li}}$ and $-a_{li}\frac{\partial}{\partial a_{li}}$. The other cases can be proved in the same way. For convenience, we denote $e_*^{(0)}(-a_{0l}\frac{\partial}{\partial a_{0l}})$, $e_*^{(l)}(-a_{li}^{-1}\frac{\partial}{\partial z_{li}})$ and $e_*^{(l)}(-a_{li}\frac{\partial}{\partial a_{li}})$ by $Y_{01}$, $Y_{l1}$ and $Y_{l2}$, respectively. It is easy to see that

$$\begin{aligned}(3.72)\quad \gamma_{\hat{K}}(P_0;P_1,\dots,P_k) &= (z_{01}+a_{01}z_{11},\dots,z_{01}+a_{01}z_{1j_1},\\ &\dots,z_{0k}+a_{0k}z_{k1},\dots,z_{0k}+a_{0k}z_{kj_k}; a_{01}a_{11},\dots,a_{01}a_{1j_1},\dots,a_{0k}a_{k1},\dots,a_{0k}a_{kj_k}),\end{aligned}$$

$$g(((\gamma_{\hat{K}})_*(Y_{01}))^{(q)}) = z_{li}g^{(q)}(-1) + g^{(q)}(0), \tag{3.73}$$

$$g(((\gamma_{\hat{K}})_*(Y_{l1}))^{(q)}) = g^{(q)}(-1), \tag{3.74}$$

$$g(((\gamma_{\hat{K}})_*(Y_{l2}))^{(q)}) = g^{(q)}(0). \tag{3.75}$$

Let

$$U_{-1,-1} = \frac{[g(0),[g(-1),g(-1)]]}{2} = -[g(-1),[g(0),g(-1)]], \tag{3.76}$$

$$U_{0,-1} = [g(0),[g(0),g(-1)] = \frac{[g(-1),[g(0),g(0)]]}{2}. \tag{3.77}$$

Also for topological vertex algebras, we fix a choice of $U_{0,0}$ in (2.25). Let $U^{(q)}_{-1,-1}$, $U^{(q)}_{0,-1}$ and $U^{(q)}_{0,0}$ be the operators on $V^{\otimes(j_1+\cdots+j_k)}$ equal to $U_{-1,-1}$, $U_{0,-1}$ and $U_{0,0}$, respectively, acting on the $q$-th tensor factor. For topological vertex algebras, the bracket $[g(Z_1),g(Z_2)]$ is $Q$-exact for any tangent vectors $Z_1,Z_2\in T_{\gamma_{\hat{K}}(P_0;P_1,\dots,P_k)}(\hat{K}(j_1+\cdots+j_k))$. The bracket $[g(Z_1),g(Z_2)]$ can be written as a linear combination of $[g^{(r)}(-1),g^{(r)}(-1)]$, $[g^{(r)}(0),g^{(r)}(-1)]$ and $[g^{(r)}(0),g^{(r)}(0)]$, $r = 1,\dots,j_1+\cdots+j_k$. For any such $Z_1$ and $Z_2$ we define an operator $U_{Z_1,Z_2}$ on $V^{\otimes(j_1+\cdots+j_k)}$ by replacing $[g^{(r)}(-1),g^{(r)}(-1)]$, $[g^{(r)}(0),g^{(r)}(-1)]$ and $[g^{(r)}(0),g^{(r)}(0)]$, $r = 1,\dots,j_1+\cdots+j_k$, in the linear expansion of $[g(Z_1),g(Z_2)]$ by $U^{(r)}_{-1,-1}$, $U^{(r)}_{0,-1}$ and $U^{(r)}_{0,0}$, respectively. Then

$$[g(Z_1),g(Z_2)] = [Q_{V^{\otimes(j_1+\cdots+j_k)}}, U_{Z_1,Z_2}] \tag{3.78}$$

by (2.25), (2.27) and (2.28).

We have

$$A^{(q)} = (z_{li}g^{(q)}(-1) + g^{(q)}(0)) \wedge g^{(q)}(-1) \wedge g^{(q)}(0). \tag{3.79}$$

where the right hand side is the skew-symmetrization of

$$(z_{li}g^{(q)}(-1) + g^{(q)}(0))g^{(q)}(-1)g^{(q)}(0).$$

It is a straightforward calculation to obtain

$$\begin{aligned}
&(z_{li}g^{(q)}(-1) + g^{(q)}(0)) \wedge g^{(q)}(-1) \wedge g^{(q)}(0)\\
&=(z_{li}g^{(q)}(-1) + g^{(q)}(0))\frac{[g^{(q)}(-1), g^{(q)}(0)]_1}{2}\\
&\quad - \frac{1}{2}g^{(q)}(0)[(z_{li}g^{(q)}(-1) + g^{(q)}(0)), g^{(q)}(-1)]\\
&\quad + \frac{1}{2}g^{(q)}(-1)[(z_{li}g^{(q)}(-1) + g^{(q)}(0)), g^{(q)}(0)]\\
&\quad - \frac{1}{3}[[g^{(q)}(-1), (z_{li}g^{(q)}(-1) + g^{(q)}(0))], g^{(q)}(0)]\\
&\quad + \frac{1}{3}[[g^{(q)}(0), (z_{li}g^{(q)}(-1) + g^{(q)}(0))], g^{(q)}(-1)].
\end{aligned} \tag{3.80}$$

Now from (3.79), (3.80), (3.76), (3.77), (3.73) — (3.75), (3.43) and (3.78) we obtain

$$\begin{aligned}
A^{(q)} &= (z_{li}g^{(q)}(-1) + g^{(q)}(0)) \wedge g^{(q)}(-1) \wedge g^{(q)}(0)\\
&=g(((\gamma_{\hat{K}})_*(Y_{01}))^{(q)})\frac{[g(((\gamma_{\hat{K}})_*(Y_{l1}))^{(q)}), g(((\gamma_{\hat{K}})_*(Y_{l2}))^{(q)})]_1}{2}\\
&\quad + \frac{1}{2}[Q_{V^{\otimes(j_1+\cdots+j_k)}}, g(((\gamma_{\hat{K}})_*(Y_{l2}))^{(q)})U^{(q)}_{(\gamma_{\hat{K}})_*(Y_{01}),(\gamma_{\hat{K}})_*(Y_{l1})}]\\
&\quad - \frac{1}{2}[Q_{V^{\otimes(j_1+\cdots+j_k)}}, g(((\gamma_{\hat{K}})_*(Y_{l1}))^{(q)})U^{(q)}_{(\gamma_{\hat{K}})_*(Y_{01}),(\gamma_{\hat{K}})_*(Y_{l2})}]\\
&\quad - \frac{1}{2}L(((\gamma_{\hat{K}})_*(Y_{l2}))^{(q)})U^{(q)}_{(\gamma_{\hat{K}})_*(Y_{01}),(\gamma_{\hat{K}})_*(Y_{l1})}\\
&\quad + \frac{1}{2}L(((\gamma_{\hat{K}})_*(Y_{l1}))^{(q)})U^{(q)}_{(\gamma_{\hat{K}})_*(Y_{01}),(\gamma_{\hat{K}})_*(Y_{l2})}\\
&\quad - U^{(q)}_{(\gamma_{\hat{K}})_*(Y_{01}),[(\gamma_{\hat{K}})_*(Y_{l1})),(\gamma_{\hat{K}})_*(Y_{l1})]}.
\end{aligned} \tag{3.81}$$

To prove the lemma in this case, we need only to show that the sum of last three terms in (3.81) composed with $\nu(\gamma_{\hat{K}}(P_0; P_1, \ldots, P_k))$ is an exact form on the leaves of the foliation described in the lemma evaluated at

$$(\gamma_{\hat{K}})_*(Y_{01}) \otimes (\gamma_{\hat{K}})_*(Y_{l1})) \otimes (\gamma_{\hat{K}})_*(Y_{l2}).$$

We define a 2-form $\beta$ on $\hat{K}(k)\times\hat{K}(j_1)\times\cdots\times\hat{K}(j_k)$ by:

$$\beta(e_*^{(0)}(X_1), e_*^{(0)}(X_2)) = 0 \tag{3.82}$$

if $X_1, X_2 \in \Gamma(T(\hat{K}(k)))$;

$$\beta(e_*^{(l)}(X_1), e_*^{(m)}(X_2)) = 0 \tag{3.83}$$

if $X_1 \in \Gamma(T(\hat{K}(j_l)))$ $X_2 \in \Gamma(T(\hat{K}(j_m)))$, $l, m = 1, \dots, k$;

$$\beta(e_*^{(0)}(X_1), e_*^{(l)}(X_2)) = \nu(\gamma_{\hat{K}}(P_0; P_1, \dots, P_k))U_{(\gamma_{\hat{K}})_*(e_*^{(0)}(X_1)),(\gamma_{\hat{K}})_*(e_*^{(l)}(X_2))} \tag{3.84}$$

if $X_1 \in \Gamma(T(\hat{K}(k)))$, $X_2 \in \Gamma(T(\hat{K}(j_l)))$, $l = 1, \dots, k$. Using the facts that $X_1, X_2 \in \Gamma(T(\hat{K}(k)\times\hat{K}(j_1)\times\cdots\times\hat{K}(j_k)))$ can always be expanded as linear combinations of those tangent vector fields occurred in (3.82), (3.83) and (3.84) and that $\beta$ should be skew-symmetric, (3.82), (3.83) and (3.84) give a well-defined form. Now we restrict $\beta$ to leaves of the foliation of $\hat{K}(k)\times\hat{K}(j_1)\times\cdots\times\hat{K}(j_k)$ given by $z_{ms} =$ constant, $a_{ms} =$ constant, $m \neq 0, l$ or $m = 0$, $s \neq l$ or $m = l$, $s \neq i$. On such a leaf, we have

$$\begin{aligned}
&(d\beta)(Y_{01}, Y_{l1}, Y_{l2})\\
=&Y_{01}\beta(Y_{l1}, Y_{l2}) - Y_{l1}\beta(Y_{01}, Y_{l2}) + Y_{l2}\beta(Y_{01}, Y_{l1})\\
&-\beta([Y_{01}, Y_{l1}], Y_{l2}) + \beta([Y_{01}, Y_{l2}], Y_{l1}) - \beta([Y_{l1}, Y_{l2}], Y_{01})\\
=&-\nu(\gamma_{\hat{K}}(P_0; P_1, \dots, P_k))L(((\gamma_{\hat{K}})_*(Y_{l1}))^{(q)})U^{(q)}_{(\gamma_{\hat{K}})_*(Y_{01}),(\gamma_{\hat{K}})_*(Y_{l2})}\\
&-\nu(\gamma_{\hat{K}}(P_0; P_1, \dots, P_k))Y_{l1}U^{(q)}_{(\gamma_{\hat{K}})_*(Y_{01}),(\gamma_{\hat{K}})_*(Y_{l2})}\\
&+\nu(\gamma_{\hat{K}}(P_0; P_1, \dots, P_k))L(((\gamma_{\hat{K}})_*(Y_{l2}))^{(q)})U^{(q)}_{(\gamma_{\hat{K}})_*(Y_{01}),(\gamma_{\hat{K}})_*(Y_{l1})}\\
&+\nu(\gamma_{\hat{K}}(P_0; P_1, \dots, P_k))Y_{l2}U^{(q)}_{(\gamma_{\hat{K}})_*(Y_{01}),(\gamma_{\hat{K}})_*(Y_{l1})}\\
&+\nu(\gamma_{\hat{K}}(P_0; P_1, \dots, P_k))U^{(q)}_{(\gamma_{\hat{K}})_*(Y_{01}),[(\gamma_{\hat{K}})_*(Y_{l1}),(\gamma_{\hat{K}})_*(Y_{l2})]}\\
=&-\nu(\gamma_{\hat{K}}(P_0; P_1, \dots, P_k))L(((\gamma_{\hat{K}})_*(Y_{l1}))^{(q)})U^{(q)}_{(\gamma_{\hat{K}})_*(Y_{01}),(\gamma_{\hat{K}})_*(Y_{l2})}\\
&+\nu(\gamma_{\hat{K}}(P_0; P_1, \dots, P_k))L(((\gamma_{\hat{K}})_*(Y_{l2}))^{(q)})U^{(q)}_{(\gamma_{\hat{K}})_*(Y_{01}),(\gamma_{\hat{K}})_*(Y_{l1})}\\
&+2\nu(\gamma_{\hat{K}}(P_0; P_1, \dots, P_k))U^{(q)}_{(\gamma_{\hat{K}})_*(Y_{01}),[(\gamma_{\hat{K}})_*(Y_{l1}),(\gamma_{\hat{K}})_*(Y_{l2})]}.
\end{aligned} \tag{3.85}$$

Thus we see that composed with $\nu(\gamma_{\hat{K}}(P_0; P_1, \dots, P_k))$ the sum of the last three term in (3.81) is equal to $-\frac{1}{2}d\beta(Y_{01}, Y_{l1}, Y_{l2})$. This proves the lemma in this case. □

Using this lemma, we can write (3.71) as a sum of terms of the following types: the composition of $\nu(\gamma_{\hat{K}}(P_0; P_1, \dots, P_k))$ with the product of the first terms of all $A^{(q)}$, $q = 1, \dots, j_1 + \cdots + j_k$; the composition of $\nu(\gamma_{\hat{K}}(P_0; P_1, \dots, P_k))$ with the product of the $Q$-exact terms of all $A^{(q)}$, $q = 1, \dots, j_1 + \cdots + j_k$; compositions of $\nu(\gamma_{\hat{K}}(P_0; P_1, \dots, P_k))$ with products of some of the $A^{(q)}$'s and the $Q$-exact terms of the other $A^{(q)}$'s; compositions of $\nu(\gamma_{\hat{K}}(P_0; P_1, \dots, P_k))$ with products of terms in the $A^{(q)}$'s such that at least one of them is the third term in some $A^{(q)}$. Recall that the left hand side of (3.60) is a sum of expressions of the form (3.71) with suitable signs. Adding terms of the first type for all expressions of the form (3.71) with the signs, we obtain (3.64). The proof for the case of strong topological vertex algebras shows that this is equal to the first term of the right hand side of (3.60). The sum of all the terms of the second type with the signs gives a $Q_{\mathcal{H}_{V,0}}$-exact operator. Using the same method used in the proof of Proposition 3.2, it is easy to show that the sum of all the terms of the third type is a sum of $Q_{\mathcal{H}_{V,0}}$-exact operators and maps from

$$(\Gamma(T(\hat{K}(k) \times \hat{K}(j_1) \times \cdots \times \hat{K}(j_k))))^{\otimes(n_0+\cdots+n_k)}$$

to

$$\mathcal{H}_{V,0}(j_1 + \cdots + j_k) \otimes \mathrm{Hol}(\hat{K}(k) \times \hat{K}(j_1) \times \cdots \times \hat{K}(j_k))$$

evaluated at $X_{ms}$, $s = 1, \dots, n_m$, $m = 0, \dots, k$, satisfying the property that when fixing some of $X_{ms}$'s and viewed only as functions of the remaining $X_{ms}$'s, these maps are holomorphic forms on $\hat{K}(k) \times \hat{K}(j_1) \times \cdots \times \hat{K}(j_k)$ and when restricted to leaves of certain pull-back coordinate foliations of $\hat{K}(k) \times \hat{K}(j_1) \times \cdots \times \hat{K}(j_k)$, these forms are exact. The sum of all terms of the fourth type, by definition and Lemma 3.4, is a sum of maps from

$$(\Gamma(T(\hat{K}(k) \times \hat{K}(j_1) \times \cdots \times \hat{K}(j_k))))^{\otimes(n_0+\cdots+n_k)}$$

to

$$\mathcal{H}_{V,0}(j_1 + \cdots + j_k) \otimes \mathrm{Hol}(\hat{K}(k) \times \hat{K}(j_1) \times \cdots \times \hat{K}(j_k))$$

evaluated at $X_{ms}$, $s = 1, \dots, n_m$, $m = 0, \dots, k$, satisfying the property that when fixing some of $X_{ms}$'s and viewed only as functions of the remaining $X_{ms}$'s, these maps are holomorphic forms on $\hat{K}(k) \times \hat{K}(j_1) \times \cdots \times \hat{K}(j_k)$ and when restricted to leaves of the coordinate foliations of $\hat{K}(k) \times \hat{K}(j_1) \times \cdots \times \hat{K}(j_k)$ described in Lemma 3.4, these forms are exact. Thus we see that the sum of all the terms of the second, third and the fourth types gives $h$ and $\alpha$.

This concludes the proof of (3.60) in the case of topological vertex algebras. The identities (3.61) and (3.62) are obvious. The proof for weak topological vertex algebras is the same as that for topological vertex algebras. □

The last property concerns the action of the symmetry groups on $\omega_j$.

**Proposition 3.5.** *Let $\sigma$ be an element of the symmetry group $S_j$ and $P$ an element of $\hat{K}(j)$. Then*

$$\sigma(\omega_j\big|_P) = \omega\big|_{\sigma(P)}. \tag{3.86}$$

*Proof.* This proposition follows from the fact that $\nu$ is a morphism of partial operad. □

## 4. The operadic formulation of (weak, strong) topological vertex algebras and the equivalence theorem

In Section 5, we will see that the meromorphic forms $\omega_j$ and their four properties give all the nice topological properties. Therefore it is natural to introduce the following notion of (weak, atrong) topological $\hat{K}$-associative algebra in terms of these meromorphic forms and the four properties:

**Definition 4.1.** A *topological $\hat{K}$-associative algebra* is a $\mathbb{Z}\times\mathbb{Z}$-graded space $V$ (graded by *weight* and *fermion number*) together with a differential $Q$ of fermion number 1 on $V$, that is, an operator $Q$ of fermion number 1 on $V$ satisfying $Q^2 = 0$, and with a $\mathrm{Hom}(V^{\otimes j}, \overline{V})$-valued holomorphic form $\omega_j$ on $\hat{K}(j)$ for every $j \in \mathbb{N}$ satisfying (3.22), Proposition 3.1, (3.29), (3.60) — (3.62) and (3.86). *Weak topological $\hat{K}$-associative algebras* and *strong topological $\hat{K}$-associative algebras* are defined defined in the same way except that the axiom (3.60) — (3.62) are replaced by the corresponding statements for weak and strong topological vertex algebras in Proposition 3.3, respectively.

We denote a (weak, strong) topological $\hat{K}$-associative algebra by $(V, Q, \omega)$ where $V$ is the underlying $\mathbb{Z} \times \mathbb{Z}$-graded vector space, $Q$ the differential on $V$ and $\omega = \{\omega_j \mid j \in \mathbb{N}\}$. Obviously, a topological $\hat{K}$-associative algebra is a weak topological $\hat{K}$-associative algebra and a strong topological $\hat{K}$-associative algebra is a topological $\hat{K}$-associative algebra.

**Proposition 4.1.** *Let $(V, Q, \{\omega_j \mid j \in \mathbb{N}\})$ be a weak topological $\hat{K}$-associative algebra. The pair $(V, \{\omega_{j,0} \mid j \in \mathbb{N}\})$ is a graded meromorphic $\hat{K}$-associative algebra.*

*Proof.* The proposition is an immediate consequence of (3.22), Proposition 3.1, (3.59) in the case $n_0 = 0$ and (3.86). □

As expected, the notion of (weak, strong) topological $\hat{K}$-associative algebra is equivalent to the notion of (weak, strong) topological vertex algebra introduced in Section 2. The precise statement is the following theorem:

**Theorem 4.2.** *The category of (weak, strong) topological $\hat{K}$-associative algebras and the category of (weak, strong) topological vertex algebras are isomorphic.*

*Proof.* We already know that the category of graded meromorphic $\hat{K}$-associative algebras and the category of graded vertex algebras are isomorphic. The results in the previous section can be summarized as saying that given a (weak, strong) topological vertex algebra, we can construct a (weak, strong) topological $\hat{K}$-associative algebra using the graded meromorphic $\hat{K}$-associative algebra corresponding to the graded vertex algebra underlying the original (weak, strong) topological vertex algebra. Thus we need only to show that given a (weak, strong) topological $\hat{K}$-associative algebra we can find operators $Q$, $g(0)$ and $g(-1)$ on the corresponding graded vertex algebra such that the corresponding graded vertex algebra together with these operators is a (weak, strong) topological vertex algebra, and to show that the procedure to obtain a (weak, strong) topological $\hat{K}$-associative algebra from a (weak, strong) topological vertex algebra and the procedure to obtain a (weak, strong) topological vertex algebra from a (weak, strong) topological $\hat{K}$-associative algebra are inverse to each other.

Given a weak topological $\hat{K}$-associative algebra $(V, Q, \omega)$, let $(V, Y, \mathbf{1})$ be the graded vertex algebra corresponding to the graded $\hat{K}$-associative algebra $(V, \{\omega_{j,0} \mid j \in \mathbb{N}\})$. We already have the differential $Q$. The operator $g(0)$ and $g(-1)$ are defined by

$$g(0) = \omega_{1,1}(-a_1 \frac{\partial}{\partial a_1})\big|_I, \tag{4.1}$$

$$g(-1) = \omega_{1,1}(-a_1^{-1} \frac{\partial}{\partial z_1})\big|_I \tag{4.2}$$

where $I$ is the identity of the partial operad $\hat{K}$. From

$$d\omega_{1,0} = Q_{\mathcal{H}_{V,0}}(\omega_{1,1}), \tag{4.3}$$

we obtain

$$\begin{aligned} -a_1 \frac{\partial}{\partial a_1} \omega_{1,0}\big|_I &= d\omega_{1,0}(-a_1 \frac{\partial}{\partial a_1})\big|_I \\ &= (Q_{\mathcal{H}_{V,0}}(\omega_{1,1}))(-a_1 \frac{\partial}{\partial a_1})\big|_I \\ &= Q\omega_{1,1}(-a_1 \frac{\partial}{\partial a_1})\big|_I + \omega_{1,1}(-a_1 \frac{\partial}{\partial a_1})Q\big|_I \\ &= [Q, g(0)]. \end{aligned} \tag{4.4}$$

But from the definition of $L(0)$ for $(V, Y, \mathbf{1})$ we have

$$-a_1 \frac{\partial}{\partial a_1} \omega_{1,0}\big|_I = \omega_{1,0}\big|_I L(0) = L(0). \tag{4.5}$$

Thus (2.15) is proved. Similarly we can prove (2.16). The formula (2.21) follows immediately from $Q_{\mathcal{H}_{V,0}}(\omega_{2,0}) = 0$. To prove (2.22) we let $P = (z, 0; 1, 1) \in \hat{K}(2)$ and from (3.59) in the case $n_0 = 0$ we obtain

$$
\begin{aligned}
&\gamma_{\mathcal{H}_{V,0}}(\omega_{2,0}\big|_P; \omega_{1,1}(-a_1\frac{\partial}{\partial a_1} + z(-a_1^{-1}\frac{\partial}{\partial z_1}))\big|_I, \omega_{1,0}\big|_I)\\
&\qquad\qquad\qquad + \gamma_{\mathcal{H}_{V,0}}(\omega_{2,0}\big|_P; \omega_{1,0}\big|_I, \omega_{1,1}(-a_2\frac{\partial}{\partial a_2})\big|_I)\\
&= \omega_{2,1}((\gamma_{\hat{K}})_*(e_*^{(1)}(-a_1\frac{\partial}{\partial a_1}) + ze_*^{(1)}(-a_1^{-1}\frac{\partial}{\partial z_1})))\big|_P + \omega_{2,1}((\gamma_{\hat{K}})_*(e_*^{(2)}(-a_2\frac{\partial}{\partial a_2})))\big|_P\\
&= \omega_{2,1}((\gamma_{\hat{K}})_*(e_*^{(1)}(-a_1\frac{\partial}{\partial a_1}) + ze_*^{(1)}(-a_1^{-1}\frac{\partial}{\partial z_1}) + e_*^{(2)}(-a_2\frac{\partial}{\partial a_2})))\big|_P\\
&= \gamma_{\mathcal{H}_{V,0}}(\omega_{1,1}(-a_1\frac{\partial}{\partial a_1})\big|_I; \omega_{2,0}\big|_P).
\end{aligned}
\tag{4.6}
$$

When both sides of (4.6) acting on $v_1 \otimes v_2$ we obtain

$$
Y((g(0) + zg(-1))v_1, z)v_2 - (-1)^{|v_1|}Y(v_1, z)g(0)v_2 = g(0)Y(v_1, z)v_2 \tag{4.7}
$$

which is exactly (2.22). The formula (2.23) can be proved similarly. The formulas (2.17) — (2.20) can also be proved similarly using $I$ instead of $P$. Thus we have proved that $(V, Y, \mathbf{1})$ together with $Q$, $g(0)$ and $g(-1)$ is a weak topological vertex algebra. If $(V, Q, \omega)$ is a topological $\hat{K}$-associative algebra, then by (3.60) and (3.62) we have

$$
\begin{aligned}
&\omega_{2,1}((\gamma_{\hat{K}})_*(e_*^{(0)}(-a_1\frac{\partial}{\partial a_1}\big|_I)), (\gamma_{\hat{K}})_*(e_*^{(1)}(-a_1\frac{\partial}{\partial a_1}\big|_I)))\\
&\qquad\qquad = \gamma_{\mathcal{H}_{V,0}}(\omega_{1,1}(-a_1\frac{\partial}{\partial a_1})\big|_I, \omega_{1,1}(-a_1\frac{\partial}{\partial a_1})\big|_I) + Q_{\mathcal{H}_{V,0}}(h).
\end{aligned}
\tag{4.8}
$$

It is easy to see that

$$
(\gamma_{\hat{K}})_*(e_*^{(0)}(-a_1\frac{\partial}{\partial a_1}\big|_I)) = (\gamma_{\hat{K}})_*(e_*^{(1)}(-a_1\frac{\partial}{\partial a_1}\big|_I)). \tag{4.9}
$$

Thus the left hand side of (4.8) is zero and we conclude that the first term in the right hand side of (4.8) is $Q_{\mathcal{H}_{V,0}}(= Q)$-exact. From the definition of $g(0)$ and $\gamma_{\mathcal{H}_{V,0}}$, we see that this ($Q$-exact) first term in the right hand side of (4.8) is equal to $(g(0))^2$. This proves that $(g(0))^2$ is $Q$-exact and thus $(V, Y, \mathbf{1})$ together with $Q$, $g(0)$ and $g(-1)$ is a topological vertex algebra. If $(V, Q, \omega)$ is a strong topological vertex algebra, then (4.8) holds with $h = 0$. Thus the same argument shows that $(g(0))^2 = 0$ and thus $(V, Y, \mathbf{1})$ together with $Q$, $g(0)$ and $g(-1)$ is a strong topological vertex algebra.

Beginning with a (weak, strong) topological vertex algebra we obtain a (weak, strong) topological $\hat{K}$-associative algebra using the construction in the previous section. From this (weak, strong) topological $\hat{K}$-associative algebra we obtain a (weak, strong) topological vertex algebra using the construction above. Theorem 2.2, the constrction of $\omega_j$ and (4.1), (4.2) imply that this (weak, strong) topological vertex algebra is the same as the original one. Beginning with a (weak, strong) topological $\hat{K}$-associative algebra we obtain a (weak, strong) topological vertex algebra. From this (weak, strong) topological vertex algebra we obtain a (weak, strong) topological $\hat{K}$-associative algebra using the construction in the previous section. Using axioms in Proposition 3.3 (3.86) and the fact that $\hat{K}$ is generated by $\hat{K}(0)$, $\hat{K}(1)$ and $\hat{K}(2)$, this (weak, strong) topological $\hat{K}$-associative algebra is the same as the original one. □

*Remark* 4.1. The above theorem in fact shows that for (weak, strong) topological vertex algebras, the forms $\omega_{j,n}$ must be of the form (3.20). Thus we have a conceptual interpretation of the definition of $\omega_{j,n}$.

A natural question is whether (weak, strong) topological vertex algebras are in fact algebras over a certain operad. The answer is affirmative for strong topological vertex algebras. The category of strong topological vertex algebras is isomorphic to the category of algebras over the differential graded partial operad of the differentiable chain complexes of $\hat{K}(j)$, $j \in \mathbb{N}$, satisfying a certain meromorphicity axiom. Since this meromorphicity axiom is much more complicated to formulate than the (equivalent) one for the forms $\omega_j$, $j \in \mathbb{N}$ and since the notion of strong topological $\hat{K}$-associative algebra is conceptually simple, we will not try to give this meromorphicity axiom and therefore we will not give the complete further reformulation of the notion of strong topological vertex algebra in this paper. But for the topological application in the next section, in the rest of the section we prove a proposition which in the case of strong topological vertex algebras says that a strong topological vertex algebra has the structure of an algebra over the differential graded partial operad of the differentiable chain complexes of $\hat{K}(j)$, $j \in \mathbb{N}$.

We need the notions of graded (partial) operad and differential graded (partial) operad. A *graded operad* is an operad in the category of $\mathbb{Z}$-graded vector spaces. A *differential graded operad* is a graded operad $\mathcal{C} = \{\mathcal{C}(j) | j \in \mathbb{N}\}$ together with an operator $\delta$ of degree $-1$ on each $\mathcal{C}(j)$ such that $\delta^2 = 0$. *Graded partial operads*, *graded partial pseudo-operads*, *differential graded partial operads* and *differential graded partial pseudo-operads* are defined in the same way. Let $(\mathcal{C}_1, \delta_1)$ and $(\mathcal{C}_2, \delta_2)$ be two differential graded operads. A *morphism* from $(\mathcal{C}_1, \delta_1)$ to $(\mathcal{C}_2, \delta_2)$ is a morphism $\psi$ from the operad $\mathcal{C}_1$ to the operad $\mathcal{C}_2$ such that $\delta_2\psi = \psi\delta_1$. *Morphisms* for differential graded partial operads and for differential graded partial pseudo-operads are defined in the same way.

We consider the family of the the differentiable chain complexes $C^S(\hat{K}(j))$, $j \in \mathbb{N}$ with coefficients in $\mathbb{C}$. The natural chain maps

(4.10)
$$\zeta : C^S(\hat{K}(k)) \otimes C^S(\hat{K}(j_1)) \otimes \cdots \otimes C^S(\hat{K}(j_k)) \to C^S(\hat{K}(k) \times \hat{K}(j_1) \times \cdots \times \hat{K}(j_k))$$

(see, for example, [Mas] for details) composed with the (partial) maps

$$(\gamma_{\hat{K}})_* : C^S(\hat{K}(k) \times \hat{K}(j_1) \times \cdots \times \hat{K}(j_k)) \to C^S(\gamma_{\hat{K}}(\hat{K}(k); \hat{K}(j_1), \ldots, \hat{K}(j_k))) \tag{4.11}$$

give (partial) chain maps

$$\gamma_{C^S(\hat{K})} : C^S(\hat{K}(k)) \otimes C^S(\hat{K}(j_1)) \otimes \cdots \otimes C^S(\hat{K}(j_k)) \to C^S(\gamma_{\hat{K}}(\hat{K}(k); \hat{K}(j_1), \ldots, \hat{K}(j_k))). \tag{4.12}$$

We have

**Proposition 4.3.** *The family $C^S(\hat{K}) = \{C^S(\hat{K}(j)) \mid j \in \mathbb{N}\}$ together with the maps $\gamma_{C^S(\hat{K})}$ and the boundary operator $\partial$ is a differetial graded partial operad.*

*Proof.* The proposition follows easily from the definition and properties of $\zeta$ (see, for example, [Mas]) and the fact that $\gamma_{\hat{K}}$ are substitution (composition) maps for the partial operad $\hat{K}$. □

Since $C^S(\hat{K})$ is a differential graded partial operad, the homology $H_*(\hat{K})$ is a graded partial operad.

Let $(V, Q, \omega)$ be a weak topological $\hat{K}$-associative algebra. We consider the endomorphism partial pseudo-operad $\mathcal{H}_{V,0} = \{\mathrm{Hom}(V^{\otimes j}, \overline{V}) \mid j \in \mathbb{N}\}$. The following proposition is obvious:

**Proposition 4.4.** *The endomorphism operad partial pseudo-operad $\mathcal{H}_{V,0}$ together with the negation of the fermion number grading and the differential $Q_{\mathcal{H}_{V,0}}$ is a differential graded partial pseudo-operad.*

Let $I^n$ be the unit $n$-cube in $\mathbb{R}^n$ and $T : I^n \to \hat{K}(j)$ a differentiable singular $n$-cube. We define

$$\phi(T) = \int_T \omega_{j,n} = \int_{I^n} T^* \omega_{j,n}. \tag{4.13}$$

It is obvious that for degenerate differentiable singular $n$-cube $T$, $\phi(T) = 0$. From the definition of $C^S(\hat{K}(j))$, we see that (4.13) can be extended to obtain a homomorphism $\phi : C^S(\hat{K}(j)) \to \mathcal{H}_{V,0}(j)$ of vector spaces for every $j \in \mathbb{N}$.

**Proposition 4.5.** *Let $(V, Q, \omega)$ be a weak topological $\hat{K}$-associative algebra. The homomorphism $\phi$ is an $S_j$-equivariant chain map from $C^S(\hat{K}(j))$ to $\mathcal{H}_{V,0}(j)$ for every $j \in \mathbb{N}$. If $(V, Q, \omega)$ is a strong topological $\hat{K}$-associative algebra, the homomorphisms $\phi$ is a morphism of differential graded partial pseudo-operads from $C^S(\hat{K})$ to $\mathcal{H}_{V,0}$. In particular, in this case, $V$ is a $C^S(\hat{K})$-algebra.*

*Proof.* We first show that $\phi$ is a chain map. Let $s$ be any differentiable singular chain in $C_n^S(\hat{K}(j))$. Then using (3.29) and the Stokes' theorem, we have

$$Q_{\mathcal{H}_{V,0}}\phi(s) = Q_{\mathcal{H}_{V,0}}\int_s \omega_{j,n} = \int_s Q_{\mathcal{H}_{V,0}}\omega_{j,n} = \int_s d\omega_{j,n} = \int_{\partial s}\omega_{j,n} = \phi(\partial s). \tag{4.14}$$

The $S_j$-equivariance of $\phi$ follows easily from (3.86): Let $s$ be any differentiable singular chain in $C_n^S(\hat{K}(j))$ and $\sigma \in S_j$. Then by (3.86)

$$\sigma(\phi(s)) = \int_s \sigma(\omega_{j,n}) = \int_{\sigma(s)} \omega_{j,n} = \phi(\sigma(s)). \tag{4.15}$$

This proves the first assertion of the proposition. To prove the second assertion, let $T_0 : I^{n_0} \to \hat{K}(k)$ and $T_l : I^{n_l} \to \hat{K}(j_l)$, $l = 1, \dots, k$ be differentiable singular cubes. For convenience, we let $j_0 = k$. For weak topological $\hat{K}$-associative algebras, we consider only those $T_l$ such that the images of $T_l$ are in $\overline{K}(j_l)$, $l = 0, \dots, l$. We denote by $[T_l]$ the elements of $C^S_{n_l}(\hat{K}(j_l))$, $l = 0, \dots, k$, containing $T_l$, $l = 0, \dots, k$, respectively. From these differentiable singular cubes we obtain another differentiable singular cube

$$T = \gamma_{\hat{K}} \circ (T_{n_0} \times \cdots \times T_{n_k}) : I^{n_0+\cdots+n_k} \to \hat{K}(j_1 + \cdots + j_k) \tag{4.16}$$

when the right hand side exists. If we denote the chain containing $T$ by $[T]$, then by definition

$$\gamma_{C^S(\hat{K})}([T_0]; [T_1], \dots, [T_k]) = [T] \tag{4.17}$$

Let $\frac{\partial}{\partial x_{li}}$, $i = 1, \dots, n_l$, $l = 0, \dots, k$, be the vector fields on $\mathbb{R}^{n_l}$ parallel to the $i$-th axis, where we use $x_{li}$ to denote the coordinates of $\mathbb{R}^{n_l}$. Then

$$\begin{aligned} \phi([T_l]) &= \int_{I^{n_l}} T_l^* \omega_{j_l,n_l} \\ &= \int_{I^{n_l}} T_l^* \omega_{j_l,n_l}(\frac{\partial}{\partial x_{l1}}, \dots, \frac{\partial}{\partial x_{ln_l}}) dx_{l1} \cdots dx_{ln_l} \\ &= \int_{I^{n_l}} \omega_{j_l,n_l}((T_l)_*(\frac{\partial}{\partial x_{l1}}), \dots, (T_l)_*(\frac{\partial}{\partial x_{ln_l}})) dx_{l1} \cdots dx_{ln_l} \end{aligned} \tag{4.18}$$

for $l = 0, \dots, k$. We have

(4.19)
$$\begin{aligned}
\phi(\gamma_{C^S(\hat{K})}([T_0]; [T_1], \dots, [T_k])) &= \phi([T]) \\
&= \int_{I^{n_0+\cdots+n_k}} T^*\omega_{j_1+\cdots+j_k, n_0+\cdots+n_k} \\
&= \int_{I^{n_0+\cdots+n_k}} T^*\omega_{j_1+\cdots+j_k, n_0+\cdots+n_k}(\frac{\partial}{\partial x_{01}}, \dots, \frac{\partial}{\partial x_{kn_k}}) dx_{01}\cdots dx_{kn_k} \\
&= \int_{I^{n_0+\cdots+n_k}} \omega_{j_1+\cdots+j_k, n_0+\cdots+n_k}(T_*(\frac{\partial}{\partial x_{01}}), \dots, T_*(\frac{\partial}{\partial x_{kn_k}})) dx_{01}\cdots dx_{kn_k} \\
&= \int_{I^{n_0+\cdots+n_k}} \omega_{j_1+\cdots+j_k, n_0+\cdots+n_k}((\gamma_{\hat{K}})_*(e_*^{(0)}((T_0)_*(\frac{\partial}{\partial x_{01}}))), \\
&\qquad \dots, (\gamma_{\hat{K}})_*(e_*^{(k)}((T_k)_*(\frac{\partial}{\partial x_{kn_k}}))))) dx_{01}\cdots dx_{kn_k}.
\end{aligned}$$

For the topological application in the next section, we use (3.60) for an arbitrary topological vertex algebra in stead of (3.59) only for strong topological vertex algebras to obtain that the right hand side of (4.19) is equal to

$$\begin{aligned}
&\int_{I^{n_0+\cdots+n_k}} \gamma_{\mathcal{H}_{V,0}}(\omega_{k,n_0}((T_0)_*(\frac{\partial}{\partial x_{01}}), \dots, (T_0)_*(\frac{\partial}{\partial x_{0n_0}})); \\
&\qquad \omega_{j_1,n_1}((T_1)_*(\frac{\partial}{\partial x_{11}}), \dots, (T_k)_*(\frac{\partial}{\partial x_{1n_1}})), \dots, \\
&\qquad \omega_{j_k,n_k}((T_k)_*(\frac{\partial}{\partial x_{k1}}), \dots, (T_k)_*(\frac{\partial}{\partial x_{kn_k}}))) dx_{01}\cdots dx_{kn_k} \\
&+ \int_{I^{n_0+\cdots+n_k}} Q_{\mathcal{H}_{V,0}}(h((\gamma_{\hat{K}})_*(e_*^{(0)}((T_0)_*(\frac{\partial}{\partial x_{01}}))), \dots, \\
&\qquad (\gamma_{\hat{K}})_*(e_*^{(k)}((T_k)_*(\frac{\partial}{\partial x_{kn_k}}))))) dx_{01}\cdots dx_{kn_k} \\
&+ \int_{I^{n_0+\cdots+n_k}} \alpha((\gamma_{\hat{K}})_*(e_*^{(0)}((T_0)_*(\frac{\partial}{\partial x_{01}}))), \dots, \\
&\qquad (\gamma_{\hat{K}})_*(e_*^{(k)}((T_k)_*(\frac{\partial}{\partial x_{kn_k}})))) dx_{01}\cdots dx_{kn_k}.
\end{aligned} \tag{4.20}$$

For convenience let

$$\begin{aligned}
\tilde{h} = \int_{I^{n_0+\cdots+n_k}} & h((\gamma_{\hat{K}})_*(e_*^{(0)}((T_0)_*(\frac{\partial}{\partial x_{01}}))), \dots, \\
&(\gamma_{\hat{K}})_*(e_*^{(k)}((T_k)_*(\frac{\partial}{\partial x_{kn_k}})))) dx_{01}\cdots dx_{kn_k},
\end{aligned} \tag{4.21}$$

$$\tilde{\alpha} = \int_{I^{n_0+\cdots+n_k}} \alpha((\gamma_{\hat{K}})_*(e_*^{(0)}((T_0)_*(\frac{\partial}{\partial x_{01}}))), \dots, (\gamma_{\hat{K}})_*(e_*^{(k)}((T_k)_*(\frac{\partial}{\partial x_{kn_k}}))))dx_{01}\cdots dx_{kn_k}. \tag{4.22}$$

Using (4.18), (4.21) and (4.22), (4.20) is equal to

$$\begin{aligned}
&\gamma_{\mathcal{H}_{V,0}}(\int_{I^{n_0}} \omega_{k,n_0}((T_0)_*(\frac{\partial}{\partial x_{01}}), \dots, (T_0)_*(\frac{\partial}{\partial x_{0n_0}}))dx_{01}\cdots dx_{0n_0};\\
&\quad \int_{I^{n_1}} \omega_{j_1,n_1}((T_1)_*(\frac{\partial}{\partial x_{11}}), \dots, (T_k)_*(\frac{\partial}{\partial x_{1n_1}}))dx_{11}\cdots dx_{1n_1}, \dots,\\
&\quad \int_{I^{n_k}} \omega_{j_k,n_k}((T_k)_*(\frac{\partial}{\partial x_{k1}}), \dots, (T_k)_*(\frac{\partial}{\partial x_{kn_k}}))dx_{k1}\cdots dx_{kn_k}) + Q_{\mathcal{H}_{V,0}}(\tilde{h}) + \tilde{\alpha}\\
&= \gamma_{\mathcal{H}_{V,0}}(\phi([T_0]); \phi([T_1]), \dots, \phi([T_k])) + Q_{\mathcal{H}_{V,0}}(\tilde{h}) + \tilde{\alpha}.
\end{aligned} \tag{4.23}$$

For strong topological $\hat{K}$-associative algebras, the last two terms of the right hand side of (4.23) are zero. Thus from (4.19), (4.20) and (4.23) $\phi$ is a morphism of differential graded partial operads. This finishes the proof. □

## 5. Cohen's and Getzler's theorem and the Gerstenhaber or Batalin-Vilkovisky algebra structure

In this section we give an application of the results obtained in the previous two sections. We first recall the definition of Batalin-Vilkovisky algebras and Gerstenhaber algebras, results on these algebras and other related concepts and results. In particular, we state Cohen's and Getzler's theorem which describes Gerstenhaber and Batalin-Vilkovisky algebras using the operads of the homologies of the little disk operad and the framed little disk operad, respectively. Then by using Getzler's theorem (or Cohen's theorem) and Proposition 4.5, we show that the cohomology of a topological vertex algebra (or of a weak topological vertex algebra) has the structure of a Batalin-Vilkovisky algebra (or of a Gerstenhaber algebra).

**Definition 5.1.** A *Batalin-Vilkovisky algebra* is a graded commutative algebra $A$ together with an operator $\Delta$ of degree $-1$ satisfying $\Delta^2 = 0$ and

$$\begin{aligned}
\Delta(abc) =& \Delta(ab)c + (-1)^{|a|}a\Delta(bc) + (-1)^{(|a|-1)|b|}b\Delta(ac)\\
&- (\Delta a)bc - (-1)^{|a|}a(\Delta b)c - (-1)^{|a|+|b|}ab(\Delta c)
\end{aligned} \tag{5.1}$$

for any homogeneous elements $a, b, c$ of $A$.

In [LZ], Batalin-Vilkovisky algebras are called "coboundary Gerstenhaber algebras." In his study of the cohomology theory of associative rings and algebras [Ger], Gerstenhaber first discovered the following algebraic structure:

**Definition 5.2.** A *Gerstenhaber algebra* is a graded commutative algebra $A$ together with a bracket $[\cdot,\cdot]: A\otimes A\to A$ such that $[A^{(n)}, A^{(m)}]\subset A^{(n+m-1)}$ (where $A^{(n)}$, $A^{(m)}$, $n, m\in\mathbb{Z}$, are homogeneous components of $A$), satisfying

$$[a,b] = -(-1)^{(|a|-1)(|b|-1)}[b,a], \tag{5.2}$$

$$[a,[b,c]] = [[a,b],c] + (-1)^{(|a|-1)(|b|-1)}[b,[a,c]], \tag{5.3}$$

$$[a,bc] = [a,b]c + (-1)^{|a|(|b|-1)}b[a,c] \tag{5.4}$$

for any homogeneous elements $a, b, c$ of $A$.

(In [Get], Gerstenhaber algebras are called "braided algebras.") In fact a Batalin-Vilkovisky algebra is a special type of Gerstenhaber algebra. We have the following proposition proved by Getzler [Get] and Penkava and Schwarz [PS]:

**Proposition 5.1 (Getzler, Penkava-Schwarz).** *A Batalin-Vilkovisky algebra is a Gersatenhaber algebra $A$ equipped with an operator $\Delta$ of degree $-1$ such that $\Delta^2 = 0$ satisfying the following relation between the bracket and $\Delta$:*

$$[a,b] = (-1)^{|a|}\Delta(ab) - (-1)^{|a|}(\Delta a)b - a(\Delta b). \tag{5.5}$$

*Furthermore in a Batalin-Vilkovisky algebra, $\Delta$ satisfies the formula*

$$\Delta[a,b] = [\Delta a, b] + (-1)^{|a|-1}[a,\Delta b]. \tag{5.6}$$

Gerstenhaber algebras and Batalin-Vilkovisky algebras can also be described using certain operads. For Gerstenhaber algebras we consider the little disk operad $\mathcal{D} = \{\mathcal{D}(j)\mid j\in\mathbb{N}\}$ of Boardman and Vogt [BoV] (see also [May]) where $\mathcal{D}(j)$ is the space of all maps from the disjoint union of $j$ copies of the unit disk $D$ to the unit disk such that on each copy of the disk they restrict to a composition of a translation and multiplication by a positive number and such that the images of different copies of the disk are disjoint. Using Künneth theorem it is easy to see that the homology of a topological operad is a graded operad. Since the little disk operad is homotopically equivalent to the family of configuration spaces $\{\mathbb{F}_j(\mathbb{C})|j\in\mathbb{N}\} = \overline{K}$, the homology of the little disk operad and the homology of this family of configuration spaces are isomorphic. In [C] Cohen proves the following theorem without stating it explicitly:

**Theorem 5.2 (Cohen).** *The category of algebras over the operad $\{H_*(\mathbb{F}_j(\mathbb{C}))|j\in\mathbb{N}\} = H_*(\overline{K})$ (or the category of algebras over the operad of the homology of the little disk operad) is equivalent to the the category of Gerstenhaber algebras.*

For Batalin-Vilkovisky algebras, we have to consider the framed little disk operad $\mathcal{P} = \{\mathcal{P}(j)\mid j\in\mathbb{N}\}$ (see [Get]) where $\mathcal{P}(j)$ is the space of all maps from the disjoint union of $j$ copies of the unit disk $D$ to the unit disk such that on each copy of the disk they restrict to a composition of a translation and multiplication by an element of $\mathbb{C}^\times$ and such that the images of different copies of the disk are disjoint. In [Get],

Getzler proves the following theorem based on Cohen's theorem above (Getzler and Jones also have another proof of Cohen's theorem):

**Theorem 5.3 (Getzler).** *The category of algebras over the operad $H_*(\mathcal{P})$ is equivalent to the category of Batalin-Vilkovisky algebras.*

Note that every element of $\mathcal{D}(j)$ is determined by the centers $z_1, \dots, z_j$ of the images of the $j$ copies of the unit disk and the $j$ positive real numbers $a_1, \dots, a_j$ multiplied to the $j$ copies of the unit disk and similarly every element of $\mathcal{P}(j)$ is determined by the centers $z_1, \dots, z_j$ of images of the $j$ copies of the unit disk and the $j$ nonzero complex numbers $a_1, \dots, a_j$ multiplied to the $j$ copies of the unit disk. Therefore $\mathcal{D}$ can be embedded in $\mathcal{P}(j)$ and $\mathcal{P}(j)$ can be embedded in $\hat{K}(j) = \mathbb{F}_j(\mathbb{C}) \times (\mathbb{C}^\times)^j$. The proof of the following proposition is a direct verification:

**Proposition 5.4.** *The little disk operad $\mathcal{D}$ is a suboperad of the framed little disk operad $\mathcal{P}$. The framed little disk operad $\mathcal{P}$ is a suboperad of the partial operad $\hat{K}$.*

If we define the composition maps for the families $C^S(\mathcal{D})$ and $C^S(\mathcal{P})$ of the differentiable chain complexes $C^S(\mathcal{D}(j))$, $j \in \mathbb{N}$, and of $C^S(\mathcal{P}(j))$, $j \in \mathbb{N}$, in the same way as that for $C^S(\hat{K})$, by Proposition 5.4 we have

**Corollary 5.5.** *The families $C^S(\mathcal{D})$ and $C^S(\mathcal{P})$ are differential graded suboperad of the differential graded partial operad $C^S(\hat{K})$.*

Now we consider a topological $\hat{K}$-associative algebra (or a weak topological $\hat{K}$-associative algebra) $(V, Q, \{\omega_j | j \in \mathbb{N}\})$. By Proposition 4.5, we have an $S_j$-equivariant chain map $\phi$ from $C^S(\hat{K}(j))$ to $\mathcal{H}_{V,0}(j)$ for every $j \in \mathbb{N}$. These maps composed with the morphism $C^S(\mathcal{P}) \to C^S(\hat{K})$ (or $C^S(\mathcal{D}) \to C^S(\hat{K})$) induced from the embedding of $\mathcal{P}$ (or of $\mathcal{D}$)) in $\hat{K}$ give $S_j$-equivariant chain maps $\phi\big|_{\mathcal{P}}$ (or $\phi\big|_{\mathcal{D}}$) from $C^S(\mathcal{P}(j))$ (or $C^S(\mathcal{D}(j)))$ to $\mathcal{H}_{V,0}(j)$, $j \in \mathbb{N}$. Except for strong topological vertex algebras, these maps are not morphisms of partial pseudo-operads. But we have

**Proposition 5.6.** *Let $(V, Q, \omega)$ be a topological $\hat{K}$-associative algebra (or a weak topological $\hat{K}$-associative algebra). Then the graded vector space $H^*(V)$ has the structure of an algebra over the operad $H_*(\mathcal{P})$ (or over the operad $H_*(\mathcal{D})$).*

*Proof.* We only prove the case for topological $\hat{K}$-associative algebras. The proof for weak topological $\hat{K}$-associative algebras is similarly. From (4.19), (4.20) and (4.23), we see that for any differentiable singular cycles $s_l$ in $C^S_{n_l}(\hat{K}(j_l))$, $l = 0, \dots, k$, we have

$$\phi(\gamma_{C^S(\hat{K})}(s_0; s_1, \dots, s_k)) = \gamma_{\mathcal{H}_{V,0}}(\phi(s_0); \phi(s_1), \dots, \phi(s_k)) + Q_{\mathcal{H}_{V,0}}(\hat{h}) + \hat{\alpha} \tag{5.7}$$

where $\hat{h}, \hat{\alpha} \in \mathcal{H}_{V,0}(j_1 + \cdots + j_k)$ and $\hat{\alpha}$ is a sum of integrals of functions over the image of $s_0 \times \cdots \times s_k$ where, when restricted to leaves of certain coordinate foliations or pull-back coordinate foliations of $\hat{K}(k)\times\hat{K}(j_1)\times\cdots\times\hat{K}(j_k)$, these functions to be integrated are exact forms evaluated at the projections of the tangent vector fields on $s_0\times\cdots\times s_k$ to the tangent space of the leaves. Since $C^S(\mathcal{P})$ is a suboperad of $C^S(\hat{K})$, for any $s_l\in C^S_{n_l}(\mathcal{P})$, $l=0,\dots,k$, we have

$$\begin{aligned}(5.8)\quad &\phi\big|_{C^S(\mathcal{P})}(\gamma_{C^S(\mathcal{P})}(s_0;s_1,\dots,s_k))\\ &\qquad=\gamma_{\mathcal{H}_{V,0}}(\phi\big|_{C^S(\mathcal{P})}(s_0);\phi\big|_{C^S(\mathcal{P})}(s_1),\dots,\phi\big|_{C^S(\mathcal{P})}(s_k))+Q_{\mathcal{H}_{V,0}}(\hat{h})+\hat{\alpha}\end{aligned}$$

where $\hat{h}$ and $\hat{\alpha}$ are the same as in (5.7). From [Get] we know that the operad $H_*(\mathcal{P})$ is generated by $H_*(\mathcal{P}(1))$ and $H_*(\mathcal{P}(2))$. Using this fact and induction, it is easy to show that we can always find cycle $s'_0,\dots,s'_k$ homologous to $s_0,\dots,s_k$, respectively, such that the cycle $s'_0\times\cdots\times s'_k$ is a sum of cycles which when intersect with leaves of any coordinate foliation or pull-back coordinate foliation of $\hat{K}(k)\times\hat{K}(j_1)\times\cdots\times\hat{K}(j_k)$ are cycles with dimensions equal to the degrees of the exact forms on these leaves given by Proposition 3.3. Thus if we use $s'_0,\dots,s'_k$ instead of $s_0,\dots,s_k$ in (5.8), the term $\hat{\alpha}$ is zero. Passing to the homology and using the fact that $s'_0,\dots,s'_k$ are homologous to $s_0,\dots,s_k$, respectively, we obtain

$$\begin{aligned}(5.9)\quad &(\phi\big|_{C^S(\mathcal{P})})_*(\gamma_{H^*(\mathcal{P})}([s_0];[s_1],\dots,[s_k]))\\ &\qquad=\gamma_{H_*(\mathcal{H}_{V,0})}((\phi\big|_{C^S(\mathcal{P})})_*([s_0]);(\phi\big|_{C^S(\mathcal{P})})_*([s_1]),\dots,(\phi\big|_{C^S(\mathcal{P})})_*([s_k]))\end{aligned}$$

where $[s_0],\dots,[s_k]$ are the homology classes containing $s_0,\dots,s_k$, respectively. Therefore $(\phi\big|_{C^S(\mathcal{P})})_*$ is a morphism of partial pseudo-operad from $H_*(\mathcal{P})$ to $H_*(\mathcal{H}_{V,0})$. It is easy to see that the graded partial pseudo-operad $H_*(\mathcal{H}_{V,0})$ is isomorphic to the endomorphism graded partial pseudo-operad $\mathcal{H}_{H^*(V),0}$ of the cohomology $H^*(V)$ of $V$. Thus we can identify $\mathcal{H}_{H^*(V),0}$ with $H_*(\mathcal{H}_{V,0})$. From (2.15) we see that elements of $H^*(V)$ in fact can be uniquely represented by elements of $V_{(0)}$. This implies that the endomorphism graded partial pseudo-operad $\mathcal{H}_{H^*(V),0}$ is in fact equal to the endomorphism graded operad $\mathcal{E}_{H^*(V)}$ and we have a morphism $(\phi\big|_{C^S(\mathcal{P})})_*$ of graded operads from $H_*(\mathcal{P})$ to $\mathcal{E}_{H^*(V)}$. Thus $H^*(V)$ is an algebra over $H_*(\mathcal{P})$. □

Combining this proposition and Theorem 4.2 with Getzler's theorem (Theorem 5.3) or Cohen's theorem (Theorem 5.2), we obtain:

**Theorem 5.7.** *Let $V$ be a topological vertex algebra (or a weak topological vertex algebra). The cohomology $H^*(V)$ of $V$ has the structure of a Batalin-Vikovisky algebra (or of a Gerstenhaber algebra).*

We now would like to derive the concrete expressions of the operator $\Delta$, the product and the bracket on $H^*(V)$. We only discuss the case for topological vertex algebras. For weak topological vertex algebra, the discussion is similar. From [Get] we see that the operator $\Delta$ is given by the image of the generator of $H_1(\mathcal{P}(1))$ under $\psi$ and the product is given by the image of the unique element of $H_0(\mathbb{F}_2(\mathbb{C})) \subset H_0(\mathcal{P}(2))$ under the morphism $\psi$. Thus from the definition of $\psi$ we see that $\Delta = g(0)$. Since any element of $\mathbb{F}_2(\mathbb{C})$ is a cycle representing the unique element of $H_0(\mathbb{F}_2(\mathbb{C}))$, we can take the cycle to be of the form $P = (z, 0)$. Let $u$, $v \in V$ and $[u]$, $[v] \in H^*(V)$ the cohomology class containing $u$, $v$, respectively. Then the product $[u][v]$ of $[u]$ and $[v]$ is $[Y(u, z)v]$ which is the cohomology class containing $Y(u, z)v \in \tilde{V}$. Since elements of $H^*(V)$ can be represented uniquely by elements of $V_{(0)}$, we can assume $u$ and $v$ have weight 0. Since $Y(u, z)v = \sum_{n\in\mathbb{Z}} u_n v z^{-n-1}$ and wt $u_n v =$ wt $u-n-1+$wt $v = -n-1$, $[u_n v] = 0$ for $n \neq -1$. Thus $[Y(u, z)v] = [u_{-1}v]$. Since the bracket can be obtained from $\Delta$ and the product, we have recovered the construction of Lian and Zuckerman completely.

Department of Mathematics, University of Pennsylvania, Philadelphia, PA 19104
*E-mail address*: yzhuang@math.upenn.edu